\documentclass{interact}
\usepackage[utf8]{inputenc}
\usepackage[english]{babel}

\usepackage{amsmath}
\usepackage{graphicx}

\usepackage{geometry}
\usepackage{color}

\usepackage{caption}
\usepackage{subcaption}
\usepackage{natbib}
\bibpunct[, ]{(}{)}{;}{a}{}{,}

\newcommand{\pD}[2]{\frac{\partial #1}{\partial #2}}

\newcommand{\psiu}{\psi_\mathrm{u}}
\newcommand{\psic}{\psi_\mathrm{c}}
\newcommand{\phiu}{\phi_\mathrm{u}}
\newcommand{\phic}{\phi_\mathrm{c}}

\begin{document}
\title{Benchmarks for infinite medium, time dependent transport problems with isotropic scattering}
\author{
\name{William Bennett and Ryan G.\ McClarren}
\affil{Department of Aerospace and Mechanical Engineering, University of Notre Dame, Notre Dame, Indiana, United States}
}
\date{\today}
\maketitle

\begin{abstract}
The widely used AZURV1 transport benchmarks package provides a suite of solutions to isotropic scattering transport problems with a variety of initial conditions \citep{ganapol}. Most of these solutions have an initial condition that is a Dirac delta function in space; as a result these benchmarks are challenging problems to use for verification tests in computer codes. Nevertheless, approximating a delta function in simulation often leads to low orders of convergence and the inability to test the convergence of high-order numerical methods. While there are examples in the literature of integration of these solutions as Green's functions for the transport operator to produce results for more easily simulated sources, they are limited in scope and briefly explained. For a sampling of initial conditions and sources, we present solutions for the uncollided and collided scalar flux to facilitate accurate testing of source treatment in numerical solvers. The solution for the uncollided scalar flux is found in analytic form for some sources. Since integrating the Green's functions is often nontrivial, discussion of integration difficulty and workarounds to find convergent integrals is included. Additionally, our uncollided solutions can be used as source terms in verification studies, in a similar way to the method of manufactured solutions.
\end{abstract}
\section{Introduction}
The AZURV1 benchmark suite, developed by \citet{ganapol}, is an indispensable verification tool in the transport community. Some of the works that rely on these benchmarks include \citep{variansyah2022population, Harel_2021, peng2021high, Heizler_2010, garrett2013comparison, seibold2014starmap, schlachter2018hyperbolicity, hauck2019filtered, heningburg2020hybrid}. The AZURV1 benchmark extends the work of \citet{monin_1956} and contains solutions for delta-function initial conditions in planar, line,  point and spherical shell shapes. These solutions can be considered Green's functions\footnote{There is contention whether one should use ``Green's function'' or ``Green function". We follow the convention recommended by \citet{wright2006green} in retaining the possessive.} for the respective geometries. 

Since running one of the AZURV1 problems  with a numerical code requires the difficult approximation of a delta function,  \citet{garrett2013comparison} present a method for integrating Ganapol's line source problem to find the exact solution for a Gaussian initial condition that resembles the line source but is more manageable for a numerical solver. This nearby problem is cited by \citep{seibold2014starmap, schlachter2018hyperbolicity, hauck2019filtered, heningburg2020hybrid}. The popularity of this solution confirms that there is interest for such solutions in the transport community. In this work we present transport solutions for a variety of initial conditions and sources to address this need. These solutions are considerably easier to handle than Green's functions for numerical codes and are, therefore, more useful for convergence studies. Additionally, we give the solution for the uncollided scalar flux for these problems. The uncollided flux can be used as the source in a computer code that solves that transport equation. This method is similar to the Method of Manufactured Solutions (MMS) \citep{salari2000code,mcclarren2008manufactured}, where a known solution is used to solve for a source term (that is typically a complicated function of space, angle, and time), which is then given to the numerical algorithm. Other researchers can use the uncollided solutions we present as the prescribed source in a verification test.

The remainder of this paper is organized as follows. In Section \ref{sec:mcmodel} we present the multiple collision approach and the original plane pulse solution from \citep{ganapol}. We then integrate this solution over square and Gaussian spatial distributions for an initial pulse and for a source that is on for a fixed time, $t_0$ in Section \ref{sec:planeGreen}. The pulsed line source solution is presented in Section \ref{sec:cyl_gaus}, as well as the integral over this pulse in a Gaussian configuration. 

\section{Uncollided-Collided Split Transport Model}\label{sec:mcmodel}
We begin with the neutral particle transport problem in an infinite medium with isotropic scattering \citep{ganapol}
\begin{equation}\label{eq:1dtransport}
    \left(\pD{}{t} + \mu\pD{}{x} + 1 \right)\psi(x,t,\mu) = \frac{c}{2}\,\phi(x,t) + \frac{1}{2}\,S(x,t),
\end{equation}
where  the angular flux is represented by $\psi(x,t,\mu)$, the scalar flux by $\phi(x,t) = \int^1_{-1}\! d\mu'\, \psi(x,t,\mu') $, $S$ is a source, and $c$ is the scattering ratio. The spatial coordinate $x$ is measured in units of mean-free path and with a particle speed of unity, $t$ measures time in units of mean-free time; $\mu \in [-1,1]$ is the cosine of the angle between a direction of travel and the $x$-axis.  Because we are in an infinite medium, we do not specify boundary conditions. We do, however, assert that the initial conditions are zero-flux, unless otherwise specified.

We can split Eq.~\eqref{eq:1dtransport} into an equation for the angular flux of particles that have not undergone a collision and an equation for the particles that have undergone a collision.  To do this we write $\psi(x,t,\mu) = \psiu(x,t,\mu) + \psic(x,t,\mu)$.

The equation for the uncollided angular flux, $\psiu$ is 
\begin{equation}\label{eq:uncol}
    \left(\pD{}{t} + \mu\pD{}{x} + 1 \right)\psiu(x,t,\mu) =   \frac{1}{2}\,S(x,t).
\end{equation}
Notice that this equation is a purely absorbing transport equation. This fact will allow us to write closed-form solutions for the uncollided scalar flux in many instances.

The equation for the collided angular flux, $\psic$, looks like the original transport equation, Eq.~\eqref{eq:1dtransport}, with the addition of a source term from the uncollided solution:
\begin{align}\label{eq:col}
    \left(\pD{}{t} + \mu\pD{}{x} + 1 \right)\psic(x,t,\mu) &=   \frac{c}{2}\int^1_{-1}\!d\mu'\left( \psic(x,t,\mu') + \psiu(x,t,\mu')\right) \nonumber \\&=   \frac{c}{2}\int^1_{-1}\!d\mu' \psic(x,t,\mu') + S_{\mathrm{u}}(x,t).
\end{align}
Here we have defined a source term for the collided equation as
\begin{equation}
    S_{\mathrm{u}}(x,t) = \frac{c}{2}\int^1_{-1}\!d\mu' \psiu(x,t,\mu') \\=
    \frac{c}{2} \phiu(x,t).
\end{equation}

A few points are in order regarding the uncollided and collided transport equations, Eqs.~\eqref{eq:uncol} and \eqref{eq:col}. Firstly, it is clear that adding these two equations together yields the original transport equation, Eq.~\eqref{eq:1dtransport}. Furthermore, we note that if the uncollided scalar flux, $\phiu$, is known then one can solve a standard transport equation with source given by $c \phiu/2$ to get the collided solution $\psic$. This means that one could use the uncollided solutions we give below as a source term in a verification test because the solution will be an approximation to $\phic$.  

Numerical methods that employ decomposition based on collisions are often more efficient as they are able to allocate computational resources separately to solve for the typically less smooth uncollided fluxes, as in \citep{ Alcouffe_1990, Hauck_2013, Walters_2017}. Our proposed source treatment is similar to these methods, except instead of solving for the uncollided flux with a more refined mesh or more angular degrees of freedom, an analytic or semi-analytic solution for the uncollided scalar flux is inserted into Eq.~\eqref{eq:col}, acting like a source term. 

\section{Planar pulse Green's functions}\label{sec:planeGreen}
For a one dimensional infinite plane pulsed source with isotropic scattering the corresponding source is $S(x,t) = \delta(x)\delta(t)$. The solution for the uncollided scalar flux is \citep{ganapol}
\begin{equation}\label{eq:pl_uncol}
    \phiu^\mathrm{pl}(x,t)= \frac{e^{-t}}{2t}\Theta(1-|\eta|),
\end{equation}
with  
\begin{equation}\label{eq:eta_1d}
    \eta \equiv \frac{x}{t},
\end{equation}
and $\Theta$ is a step function that returns unity for positive arguments and zero otherwise. The solution for the collided flux is,
\begin{equation}\label{eq:pl_col}
    \phic^\mathrm{pl}(x,t) = c\left(\frac{e^{-t}}{8\pi}\left(1-\eta^2\right)\int^\pi_0\!du\, \mathrm{sec}^2\left(\frac{u}{2}\right)\mathrm{Re}\left[\xi^2e^{\frac{ct}{2}(1-\eta^2)\xi}\right]\right)\Theta(1-|\eta|),
\end{equation}
where the complex valued function $\xi$ is ,
\begin{equation}\label{eq:xi}
    \xi(u,\eta) = \frac{\log q + i\, u}{\eta + i \,\mathrm{tan}(\frac{u}{2})}
\end{equation}
and,
\begin{equation}\label{eq:q}
    q = \frac{1 + \eta}{1 - \eta}.
\end{equation}
Equations~\eqref{eq:pl_uncol} and \eqref{eq:pl_col} will be treated as Green's function kernels and integrated to find solutions for a variety of  sources and initial conditions. The general form for this integration is:
\begin{equation}\label{eq:general_greens}
    \phi_{\mathrm{j}}(x,t) = \int^{t}_0\!d\tau\, \int_{-\infty}^{\infty} \!d s\, S(s,\tau)\, \phi_{j}^\mathrm{pl}(x-s, t-\tau),
\end{equation}
here $S$ is an arbitrary source, and the subscript $j = $ u or c denotes the uncollided or collided scalar flux. 

\begin{figure}
     \centering
     \begin{subfigure}[b]{0.3\textwidth}
         \centering
         \includegraphics[width=\textwidth]{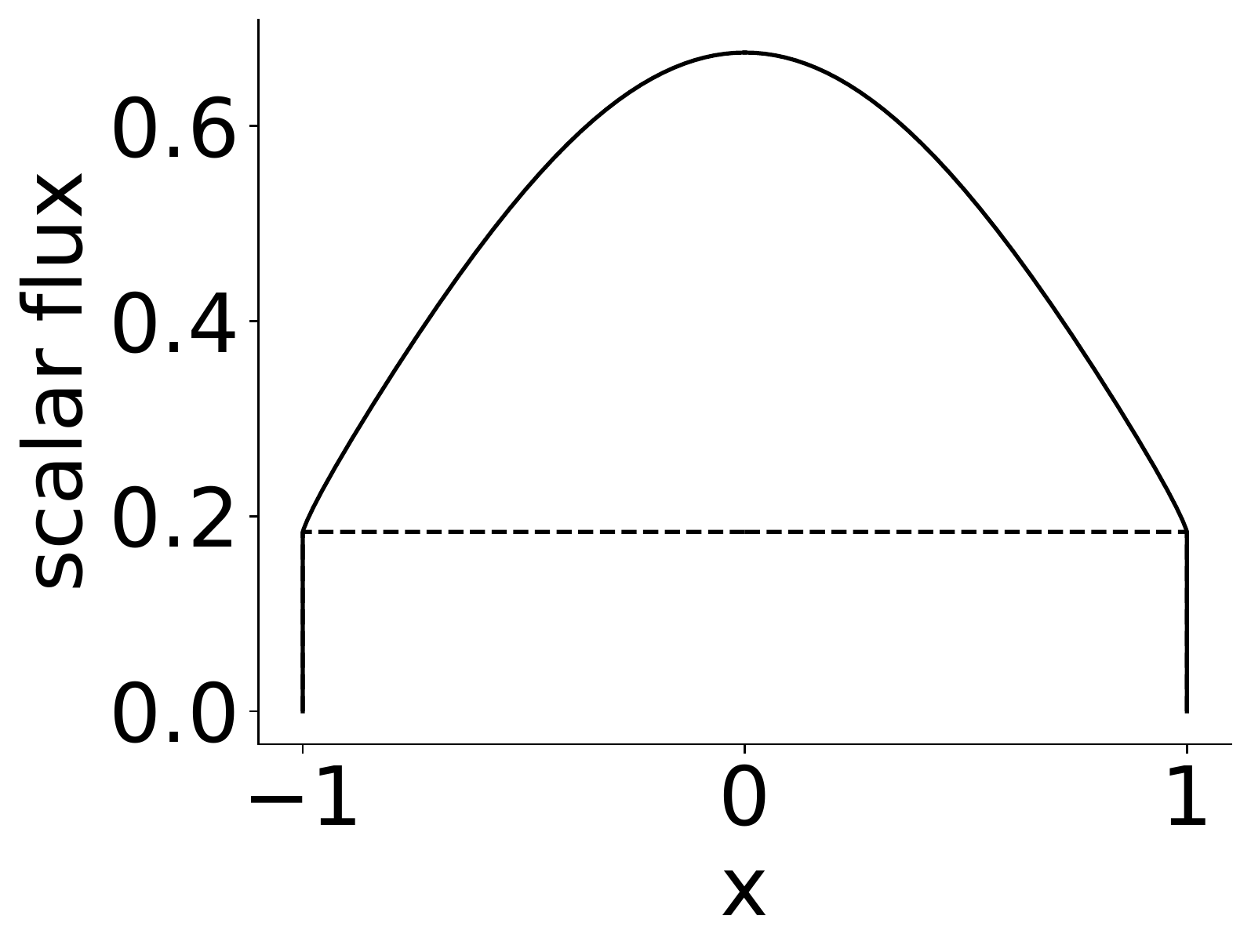}
         \caption{$t=1$}
         \label{fig:pl_IC_1}
     \end{subfigure}
     \hfill
     \begin{subfigure}[b]{0.3\textwidth}
         \centering
         \includegraphics[width=\textwidth]{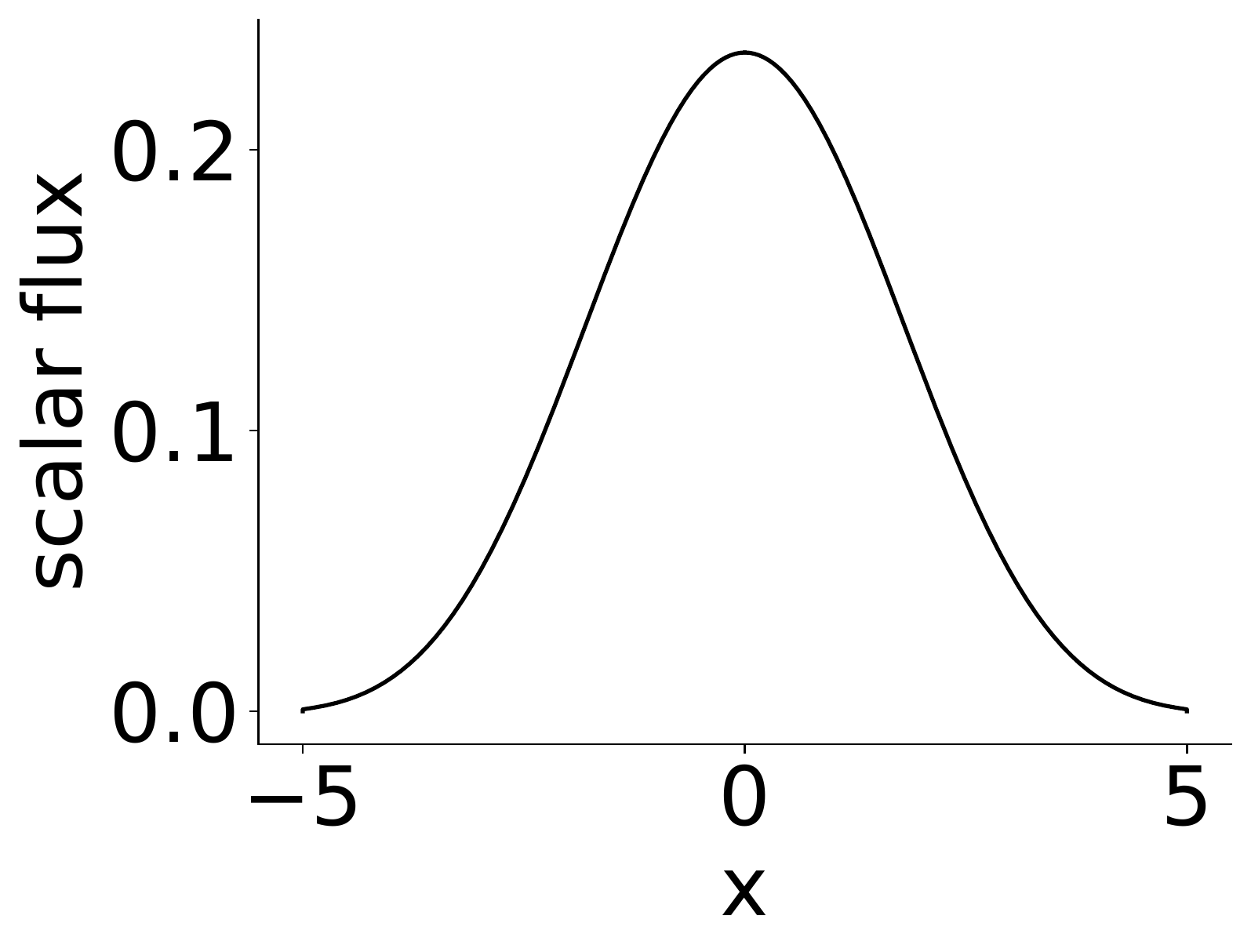}
         \caption{$t=5$}
         \label{fig:pl_IC_5}
     \end{subfigure}
     \hfill
     \begin{subfigure}[b]{0.3\textwidth}
         \centering
         \includegraphics[width=\textwidth]{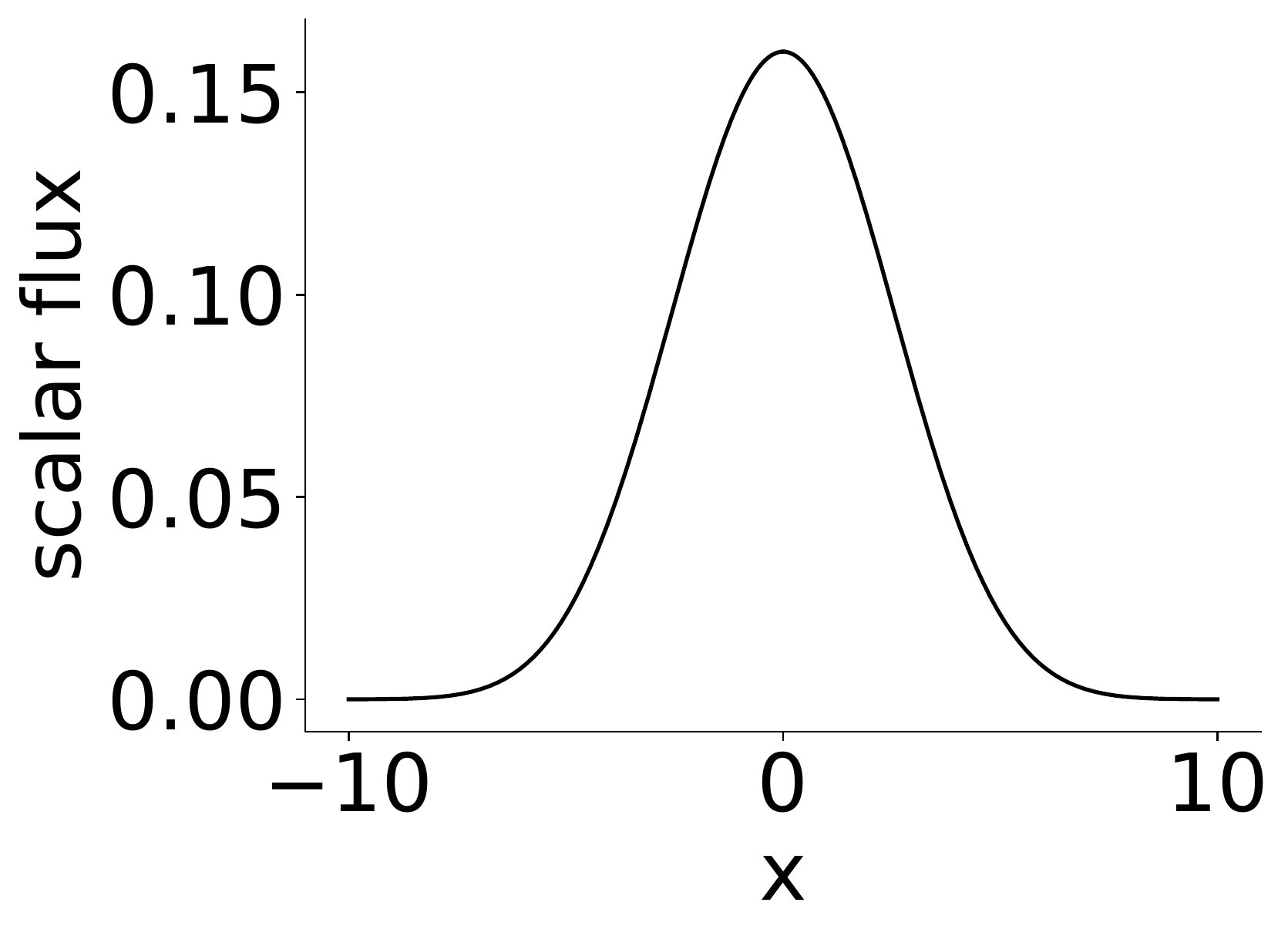}
         \caption{$t=10$}
         \label{fig:pl_IC_10}
     \end{subfigure}
        \caption{Plane pulse scalar flux solutions, $\phiu^\mathrm{pl} + \phic^\mathrm{pl}$, for $c=1$ at several times; panel (a) also contains the uncollided scalar flux, $\phiu^\mathrm{pl}$, denoted by a dashed line.}
        \label{fig:pl_IC}
\end{figure}

The total scalar flux ($\phiu + \phic$) and the uncollided scalar flux for this problem are shown in Figure \ref{fig:pl_IC}. Because the uncollided solution decays exponentially in time and is effectively zero on the scale of the plots, we do not show the uncollided solution in the later time panels. We also point out that at $t=1$ the presence of the wavefront is noticeable at $x=1$; this feature also decays exponentially and is hardly noticeable in the $t=5$ panel at $x=5$. This wavefront is one of the features in the AZURV1 solution that can make it challenging to use this benchmark in convergence studies for high-order numerical methods. 
\subsection{Square pulse}\label{sec:sqic}
The square pulse integrates the solution from the plane pulse over a finite spatial range. We consider a square pulse of width $x_0$ and magnitude one centered on the origin,  
\begin{equation}\label{eq:sqic}
    S(x,t) = \Theta(x_0 - |x|)\delta(t).
\end{equation}
This source can also be written as initial condition $\phi(x,t=0,\mu) = \Theta(x_0 - |x|)$. With Eq.~\eqref{eq:sqic} as the source, Eq.~\eqref{eq:general_greens} gives the uncollided solution:
\begin{equation}\label{eq:square_IC_1}
    \phiu^\mathrm{sp}(x,t) = \int^{x_0}_{-x_0}\!ds\,\frac{e^{-t}}{2t}\Theta\left(1-|\eta'|\right),
\end{equation}
where we have defined  $\eta'$ to contain the integration variable,
\begin{equation}\label{eq:dummy_eta_1d}
    \eta' = \frac{x-s}{t}.
\end{equation}
The Dirac delta function in the source removes the integration over $\tau$. With the integration limits changed to $[-x_0,x_0]$, the step function in the source is always unity. 
Evaluating the integral in Eq.~\eqref{eq:square_IC_1} requires considering all of the possible relationships of the parameters $x,t$ and $x_0$. With these cases considered and the integral evaluated, the uncollided solution can be written as a piecewise function:
\begin{equation}\label{eq:sq_ic_cases}
    \phiu^\mathrm{sp}(x,t) = 
    \begin{cases}
     0 & |x| - t >   x_0\\
    x_0 \frac{e^{-t}}{t} &   t > x_0 \: \& \:   x_0-t \leq x \leq t - x_0  \\
    e^{-t} & t \leq x_0 \: \& \:   t-x_0 \leq x \leq x_0-t\\
    \frac{e^{-t} (t+x+x_0)}{2 t} & -t-x_0 < x < t + x_0 \:\&\: x_0+x \leq t \leq x_0 -x \\
    \frac{e^{-t} (t-x+x_0)}{2 t} &  -t-x_0 < x < t + x_0 \: \& \: x_0-x \leq t \leq x_0 + x
    \end{cases}
\end{equation}
Eq.~\eqref{eq:general_greens} with a square pulse as the source and the plane pulse collided solution as the kernel gives,
\begin{equation}\label{eq:sq_ic_collided}
    \phic^\mathrm{sp}(x,t) = \int^{x_0}_{-x_0}\!ds\,\int^\pi_0\!du\,
   F_1(x,s,t,u)\,\Theta\left(1-|\eta'|\right),
\end{equation}
with identical simplifications to the integrals as with the uncollided case. The integrand function is defined as 
\begin{equation}\label{eq:F1sqp}
    F_1(x,s,t,u) =  \frac{c\,e^{-t}}{8\pi}\left(1-\eta'^2\right) \mathrm{sec}^2\left(\frac{u}{2}\right)\mathrm{Re}\left[\xi^2e^{\frac{ct}{2}(1-\eta'^2)\xi}\right],
\end{equation}
with 
$\xi$ given by Eq.~\eqref{eq:xi}. We have not determined a simple, closed form for the integrals in Eq.~\eqref{eq:sq_ic_collided}, though we note that the integrand appears to be a well-behaved function, and we have had no trouble performing the integration. 

\begin{figure}
     \centering
     \begin{subfigure}[b]{0.3\textwidth}
         \centering
         \includegraphics[width=\textwidth]{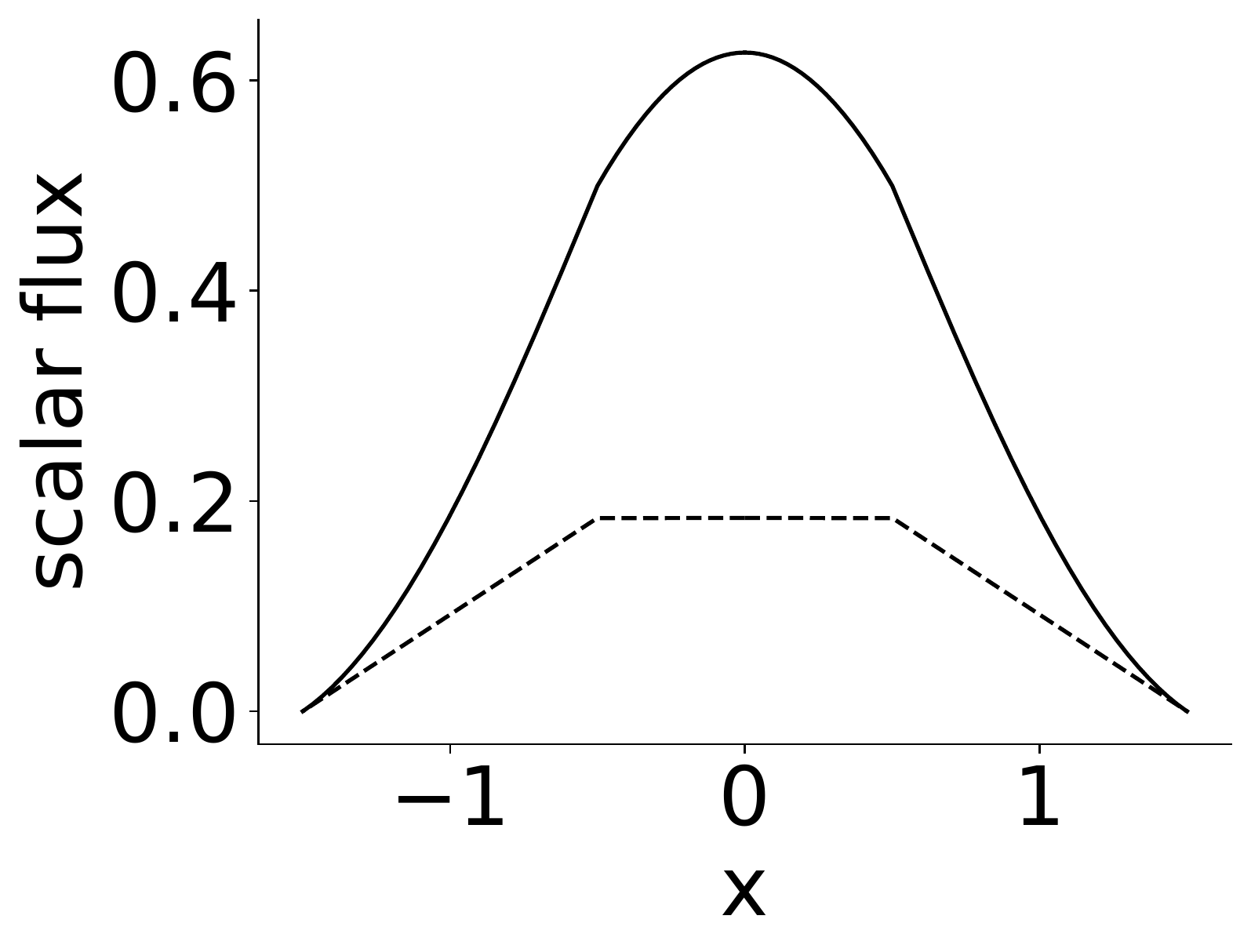}
         \caption{$t=1$}
         \label{fig:sq_IC_1}
     \end{subfigure}
     \hfill
     \begin{subfigure}[b]{0.3\textwidth}
         \centering
         \includegraphics[width=\textwidth]{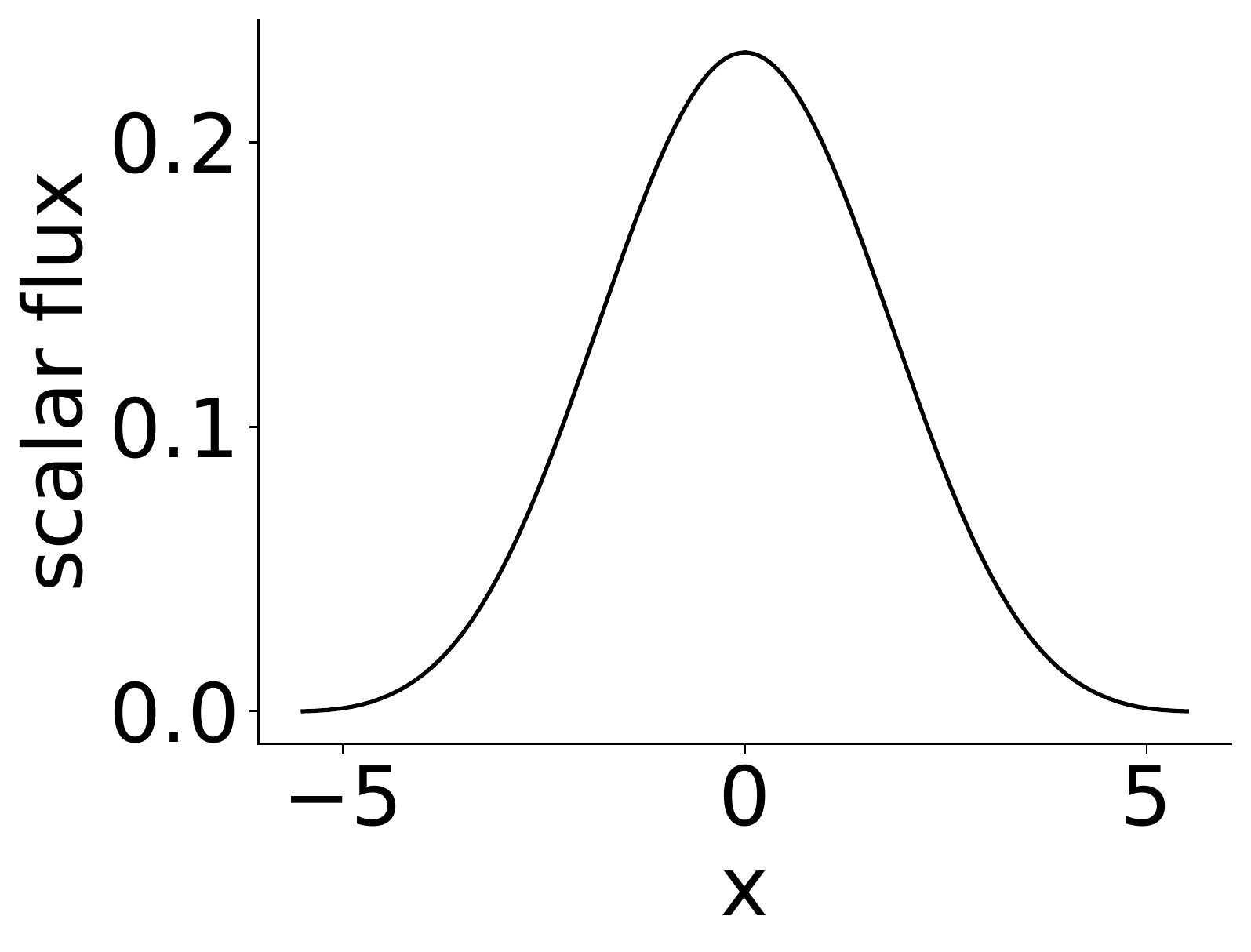}
         \caption{$t=5$}
         \label{fig:sq_IC_5}
     \end{subfigure}
     \hfill
     \begin{subfigure}[b]{0.3\textwidth}
         \centering
         \includegraphics[width=\textwidth]{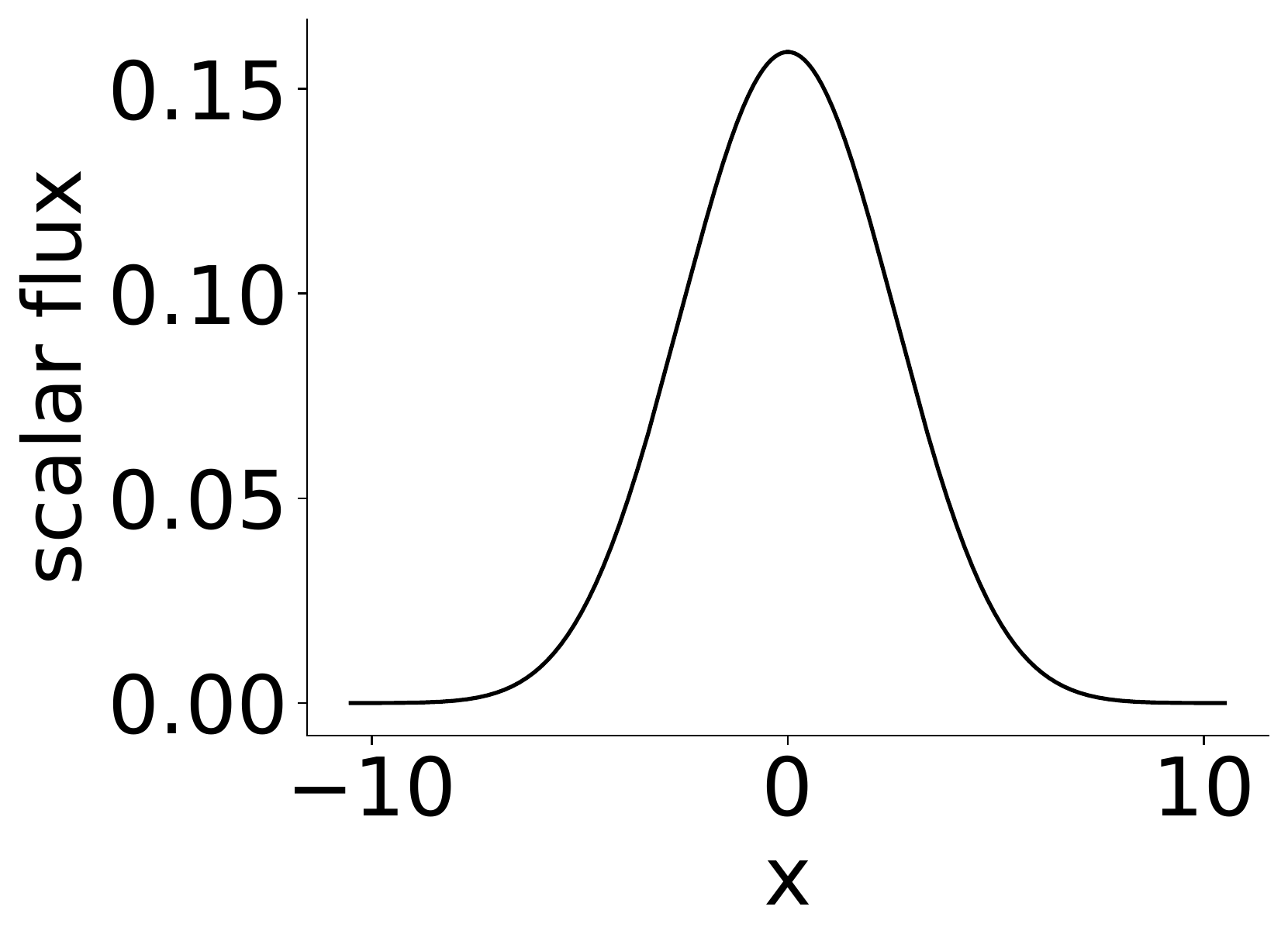}
         \caption{$t=10$}
         \label{fig:sq_IC_10}
     \end{subfigure}
        \caption{Square pulse scalar flux solutions, $\phiu^\mathrm{sp} + \phic^\mathrm{sp}$, for $c=1$ and $x_0=0.5$ at several times; panel (a) also contains the uncollided scalar flux, $\phiu^\mathrm{sp}$, denoted by a dashed line.}
        \label{fig:sq_IC}
\end{figure}

The solutions for this problem are shown in Figure \ref{fig:sq_IC} for times $t=1, 5,$ and $10$ when the source width parameter is $x_0 = 0.5$. As in the plane pulse case we also show the uncollided solution at $t=1$. Though the uncollided solution is still exponentially decaying, we notice that the uncollided solution approaches zero linearly, and is continuous, but nonsmooth at $|x|=x_0.$
\subsection{Gaussian pulse}\label{sec:gsic}
Next, we consider an initial pulse with a Gaussian spatial profile with standard deviation $\sigma$ centered on the origin,
\begin{equation}\label{eq:gp}
    S(x,t) = \exp\left({\frac{-x^2}{\sigma^2}}\right)\,\delta(t).
\end{equation}
As in Section \ref{sec:sqic}, the integration over time simplifies and the uncollided solution from Eq.~\eqref{eq:general_greens} becomes,
\begin{equation}\label{eq:gauss_uncol_1}
    \phiu^\mathrm{gp}(x,t) = \int^{\infty}_{-\infty}\!ds\,\frac{e^{-t}}{2t}\, \exp\left({\frac{-s^2}{\sigma^2}}\right)\Theta\left(1-|\eta'|\right).
\end{equation}
Equation~\eqref{eq:gauss_uncol_1} may be solved analytically. The step function defined in the Green's kernel changes the integration limits to $x-t$ and $x + t$ and the solution is 
\begin{equation}\label{eq:gaussian_pulse_uncollided}
    \phiu^\mathrm{gs}(x,t) =  \sigma\,\sqrt{\pi }\, e^{-t} \,\frac{\text{erf}\left( \frac{t-x}{\sigma}\right)+\text{erf}\left(\frac{t+x}{\sigma}\right)}{4 t}.
\end{equation}
Unlike the solution for a plane pulse or a square pulse, the uncollided angular flux induced by Eq.~\eqref{eq:gp} is smooth, which has implications for the convergence of numerical solvers. 

The expression for the collided flux in this configuration is very similar to Eq.~\eqref{eq:sq_ic_collided}, but with a different source term and integration limits,
\begin{equation}\label{eq:gauss_pulse_collided_1}
    \phic^\mathrm{gp}(x,t) = \int^{\infty}_{-\infty}\!ds\,\int^\pi_0\!du\,
    \exp\left({\frac{-s^2}{\sigma^2}}\right)\,F_1(x,s,t,u)\,\Theta\left(1-|\eta'|\right),
\end{equation}
where $F_1$ is given by Eq.~\eqref{eq:F1sqp} and $\eta'$ by Eq.~\eqref{eq:dummy_eta_1d}. In this case, finding the effective integration limits is simple. Solving $|\eta'|=1$ for $s$ gives the effective integration limits,
\begin{equation}\label{eq:gauss_pulse_collided}
    \phic^\mathrm{gp}(x,t) = \int^{x+t}_{x-t}\!ds\,\int^\pi_0\!du\,
    \exp\left({\frac{-s^2}{\sigma^2}}\right)\,F_1(x,s,t,u).
\end{equation}

\begin{figure}
     \centering
     \begin{subfigure}[b]{0.3\textwidth}
         \centering
         \includegraphics[width=\textwidth]{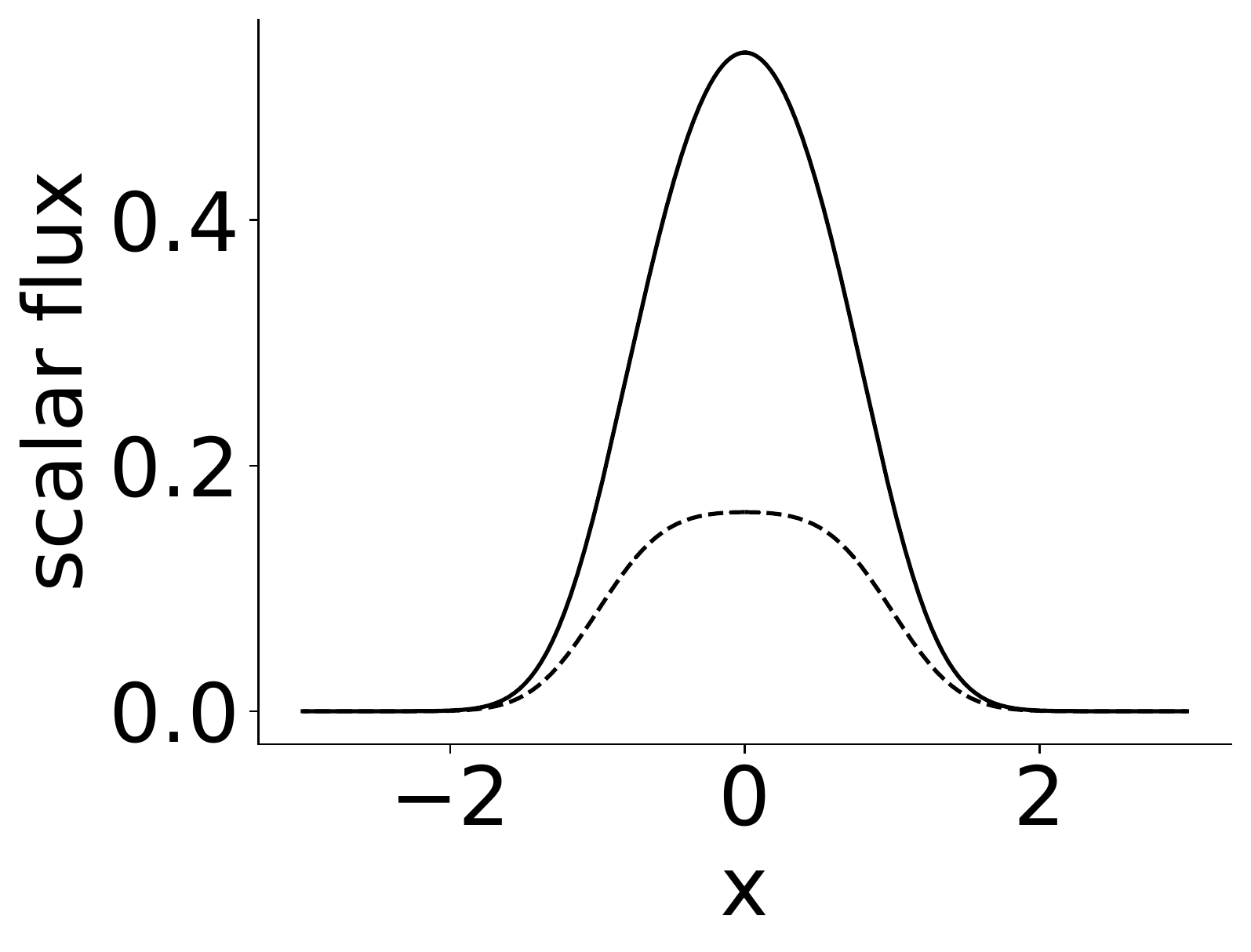}
         \caption{$t=1$}
         \label{fig:gs_IC_1}
     \end{subfigure}
     \hfill
     \begin{subfigure}[b]{0.3\textwidth}
         \centering
         \includegraphics[width=\textwidth]{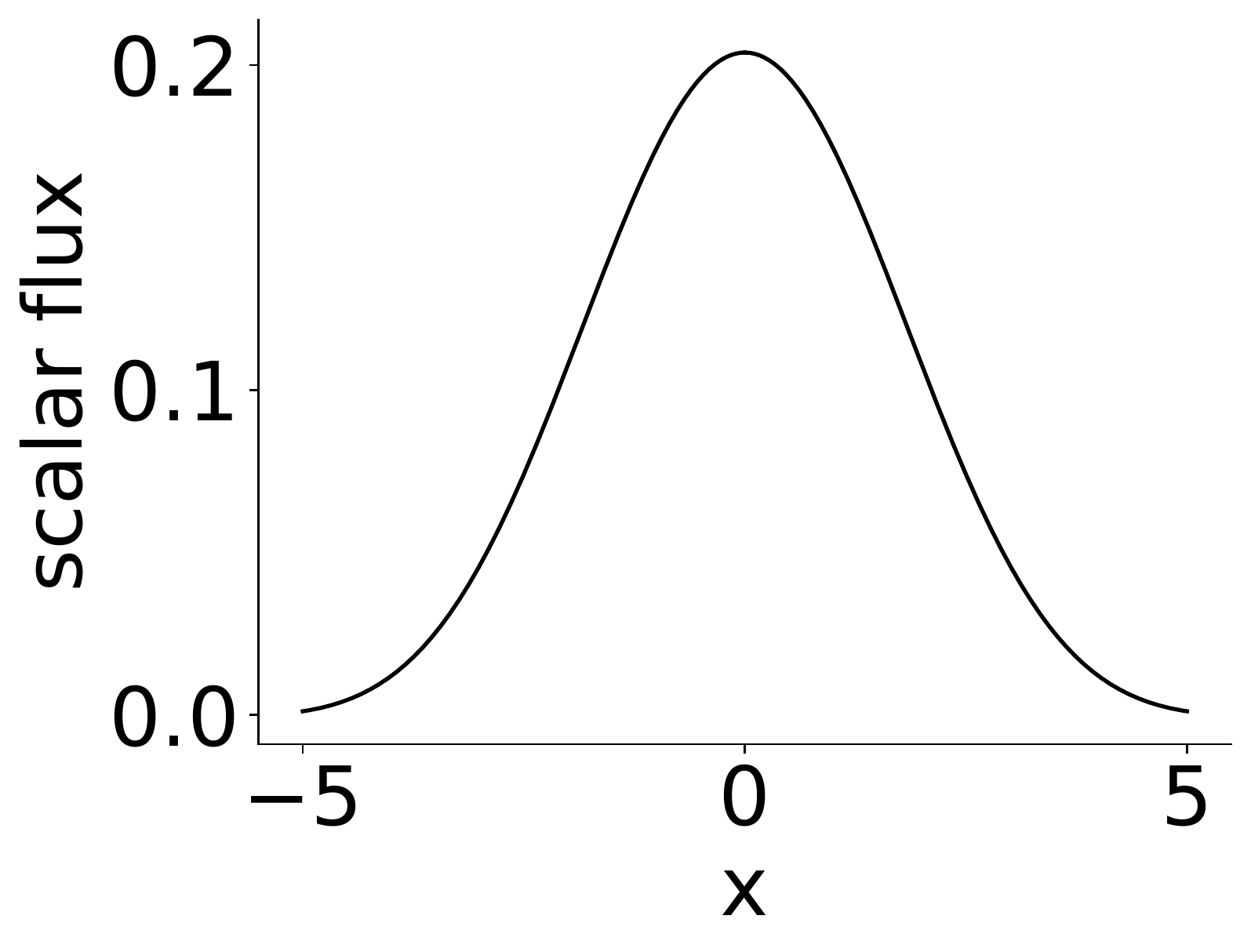}
         \caption{$t=5$}
         \label{fig:gs_IC_5}
     \end{subfigure}
     \hfill
     \begin{subfigure}[b]{0.3\textwidth}
         \centering
         \includegraphics[width=\textwidth]{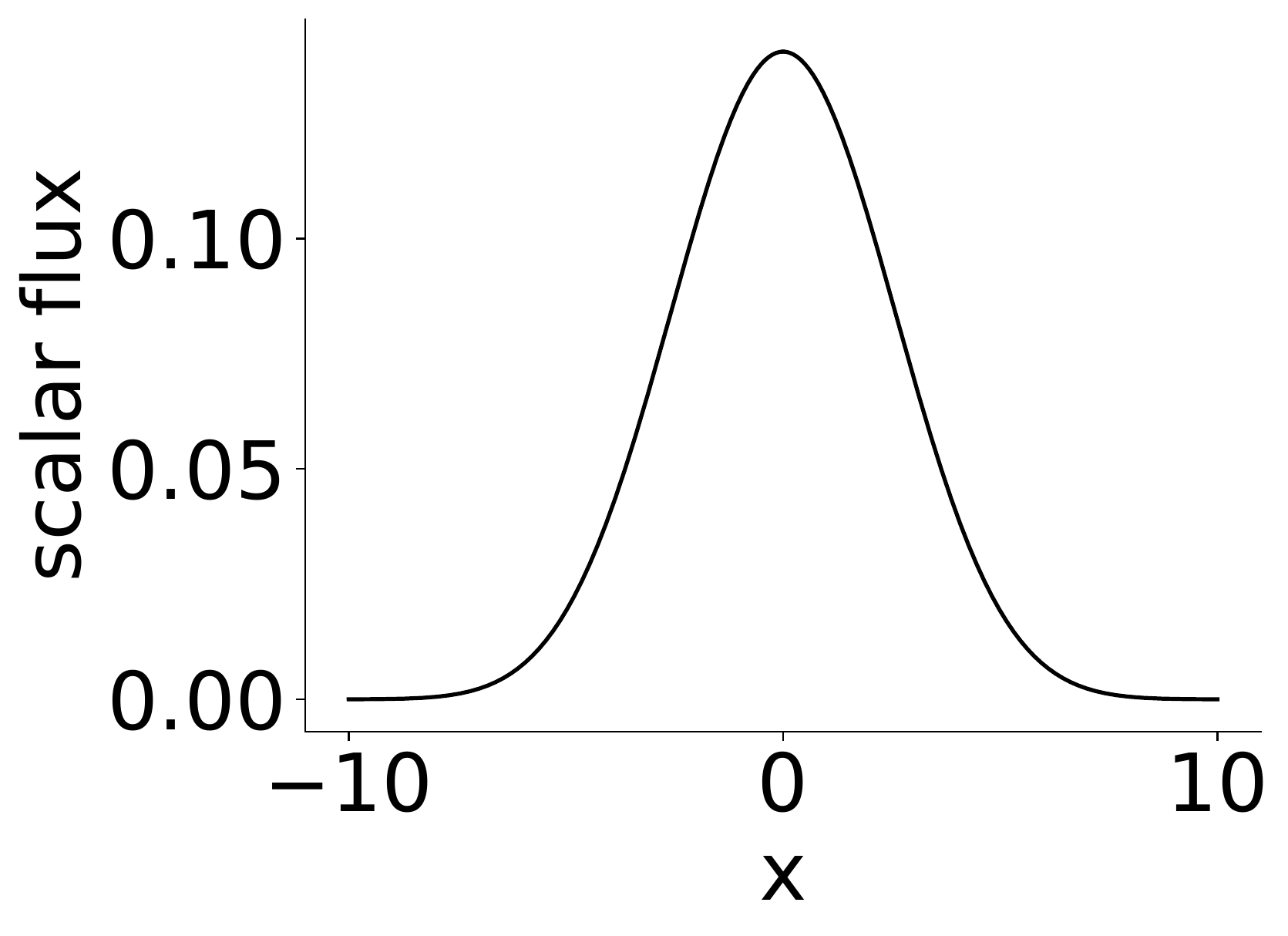}
         \caption{$t=10$}
         \label{fig:gs_IC_10}
     \end{subfigure}
        \caption{Gaussian pulse scalar flux solutions, $\phiu^\mathrm{gp} + \phic^\mathrm{gp}$, for $c=1$ and $\sigma=0.5$ at several times; panel (a) also contains the uncollided scalar flux, $\phiu^\mathrm{gp}$, denoted by a dashed line.}
        \label{fig:gs_IC}
\end{figure}

Figure \ref{fig:gs_IC} shows the solution to this problem at times $t=1,5$ and $10$. In this case all of the solutions, including the uncollided solutions, are smooth. 
\subsection{Square source}\label{sec:sqs}
We now consider sources that are nonzero for a finite length of time. The source  can be understood as superpositions of pulses, like those presented in Sections \ref{sec:sqic} and \ref{sec:gsic} from $t=0$ to $t = t_0$. This method uncovers difficulties in the Green's kernel that were overlooked in the construction of solutions for pulses. Namely, the expression $\frac{e^{-t}}{2t^n}$, where $n$ is any integer, that appears in the uncollided and collided kernels, is singular as $t$ approaches zero and the step function $\Theta(1-|\eta'|)$ behaves more erratically. This erratic behavior causes integrands to be poorly behaved in numerical integration since the effective domain changes wildly which can cause quadrature points to be wasted on regions where the integrand is zero and miss important features. 

We first consider a  source of width $x_0$ and magnitude one turned off at $t=t_0$ centered on the origin, 
\begin{equation}\label{eq:sqs}
    S(x,t) = \Theta(x_0 - |x|)\,\Theta(t_0-t).
\end{equation}
Using this source in Eq.~\eqref{eq:general_greens} we arrive at the following integral to define the uncollided square source solution
\begin{equation}\label{eq:sq_source_uncollided_1}
    \phiu^\mathrm{ss}(x,t) = \int^{\mathrm{min}(t,t_0)}_{0}\!d\tau\,\int^{x_0}_{-x_0}\!ds\,\frac{e^{-(t-\tau)}}{2(t-\tau)}\, \Theta\left(1-|\eta''|\right),
\end{equation}
with  $\eta''$ that contains the time integration variable,
\begin{equation}\label{eq:dummy_eta_1d_time}
    \eta'' = \frac{x-s}{t-\tau}.   
\end{equation}
As in Section \ref{sec:gsic}, the spatial step function in the source has been absorbed into the integration limit over $s$. The time dependent step function in the source is always one for the integration limits of Eq.~\eqref{eq:sq_source_uncollided_1}. We note that the solution for the inner integral over $s$ of  Eq.~\eqref{eq:sq_source_uncollided_1} is  Eq.~\eqref{eq:sq_ic_cases} where $t$ is replaced with $t-\tau$. 
We have,
\begin{equation}\label{eq:sq_s_integral_over_sq_p}
     \phiu^\mathrm{ss}(x,t) = \int^{\mathrm{min}(t,t_0)}_{0}\!d\tau\, \phiu^\mathrm{sp}(x,t-\tau).
\end{equation}
Solving Eq.~\eqref{eq:sq_s_integral_over_sq_p} is an exercise of accounting for all of the possible values for $\tau$. For example, the first case in the square pulse solution, the scalar flux is zero if $|x|-t>x_0$. Solving $|x|-(t-\tau)>x_0$ for $\tau$ tells us that the square source solution is zero if $\tau > t + x_0 -|x|$. Considering all of the cases in Eq.~\eqref{eq:sq_ic_cases} allows us to come to an analytic solution for the uncollided flux from a square source,
\begin{equation}
    \phiu^\mathrm{sp}(x,t) = \left[-x_0\mathrm{E_i}(\tau-t)\right]\bigg{|}_0^b + \frac{1}{2} \left[(| x| -x_0) \text{Ei}(\tau -t)+e^{\tau -t}\right]\bigg{|}_b^c + \left[e^{-(t-\tau)}\right]\bigg{|}_c^d.
\end{equation}
Where the evaluation intervals of $\tau$ are defined by
\begin{equation}
    b = \left[\mathrm{min}\left(d, t - |x| - x_0\right)\right]_+,
\end{equation}
\begin{equation}
    c = \left[\mathrm{min}\left(d, t + |x| - x_0\right)\right]_+,
\end{equation}
\begin{equation}
    d = \left[\mathrm{min}\left(t_0, t, t - |x| + x_0\right)\right]_+,
\end{equation}
where $[\cdot]_+$ returns the positive part of its argument, and
$\mathrm{E_i}$ is the exponential integral.

The expression for the collided solution takes the form of an integral over Eq.~\eqref{eq:sq_ic_collided}: 
\begin{equation}\label{eq:integral_square_source}
    \phic^\mathrm{ss}(x,t) = \int^{\mathrm{min}(t,t_0)}_0\!d\tau\,\int^{x_0}_{-x_0}\!ds\,
   F_2(x,s,t,\tau,u)\,\Theta\left(1-|\eta''|\right),
\end{equation}
where the integrand $F_2$ is slightly different from $F_1$  from the square pulse case,
\begin{equation}
    F_2(x,s,t,\tau,u) =  \frac{ce^{-(t-\tau)}}{8\pi}\left(1-\eta''^2\right) \int^\pi_0\!du\,\mathrm{sec}^2\left(\frac{u}{2}\right)\mathrm{Re}\left[\xi^2e^{\frac{c(t-\tau) 
    }{2}(1-\eta''^2)\xi}\right],
\end{equation}
with $\eta''$ given by Eq.~\eqref{eq:dummy_eta_1d_time}.

While we were able to avoid singularity difficulties by integrating analytically in finding the solution for the uncollided case in this configuration, the collided case must be integrated numerically. Due to the behavior of the step function, the effective integration domain of Eq.~\eqref{eq:integral_square_source} varies drastically with $\tau$. Therefore, the integral is not well-suited to numerical integration. Switching the order of integration and merging the step function with the integration limits over $\tau$ gives
\begin{equation}\label{eq:integrand_square_switched}
    \phic^\mathrm{ss}(x,t) = \int^{x_0}_{-x_0}\!ds\,\int^{\mathrm{min}(t,t-|x-s|)_+}_0\!d\tau\,
   F_2(x,s,t,\tau,u),
\end{equation}
 and allows us to cast the integral in a form that we have found to converge at an reasonable rate. 

\begin{figure}
     \centering
     \begin{subfigure}[b]{0.3\textwidth}
         \centering
         \includegraphics[width=\textwidth]{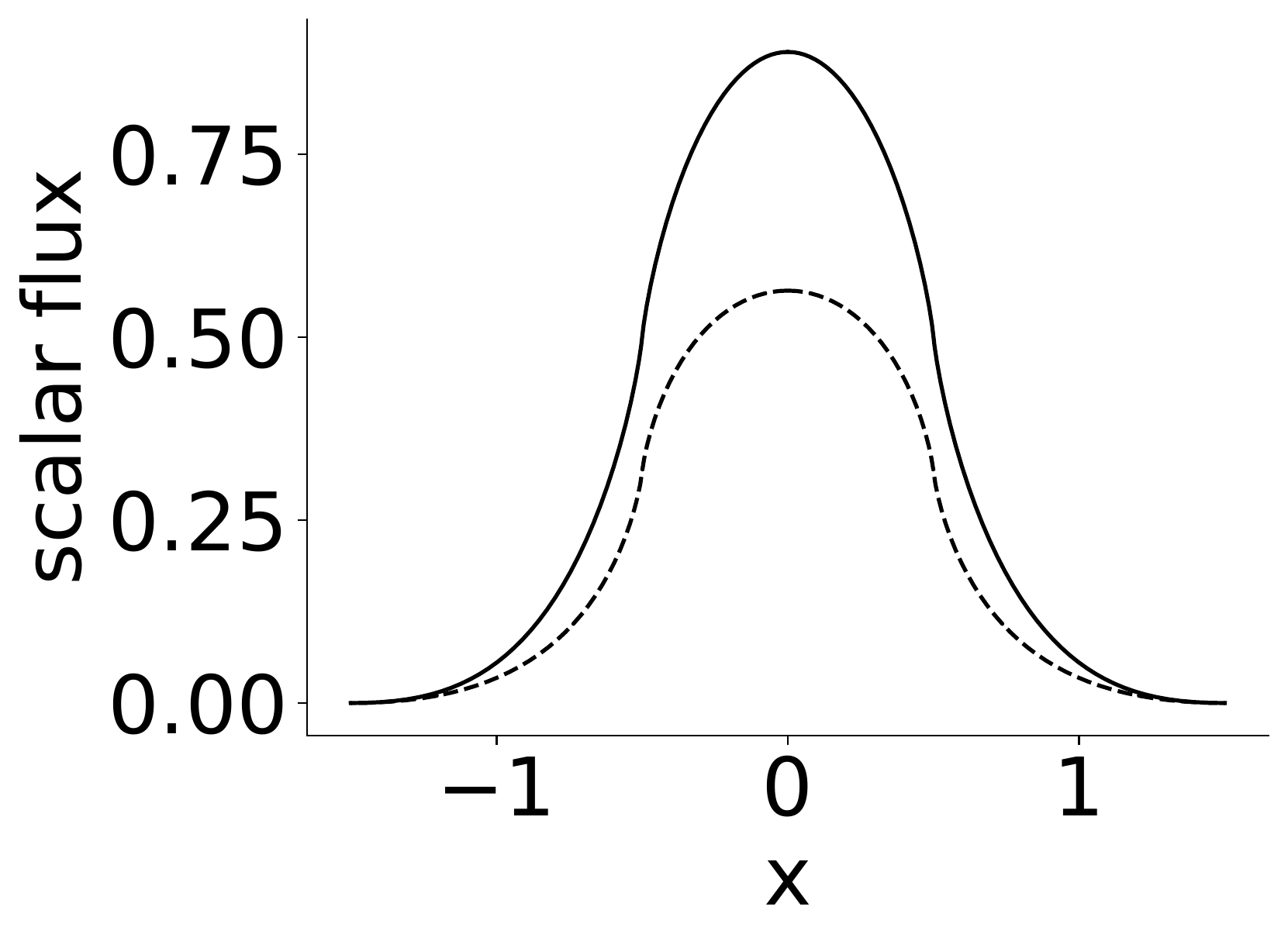}
         \caption{$t=1$}
         \label{fig:sq_s_1}
     \end{subfigure}
     \hfill
     \begin{subfigure}[b]{0.3\textwidth}
         \centering
         \includegraphics[width=\textwidth]{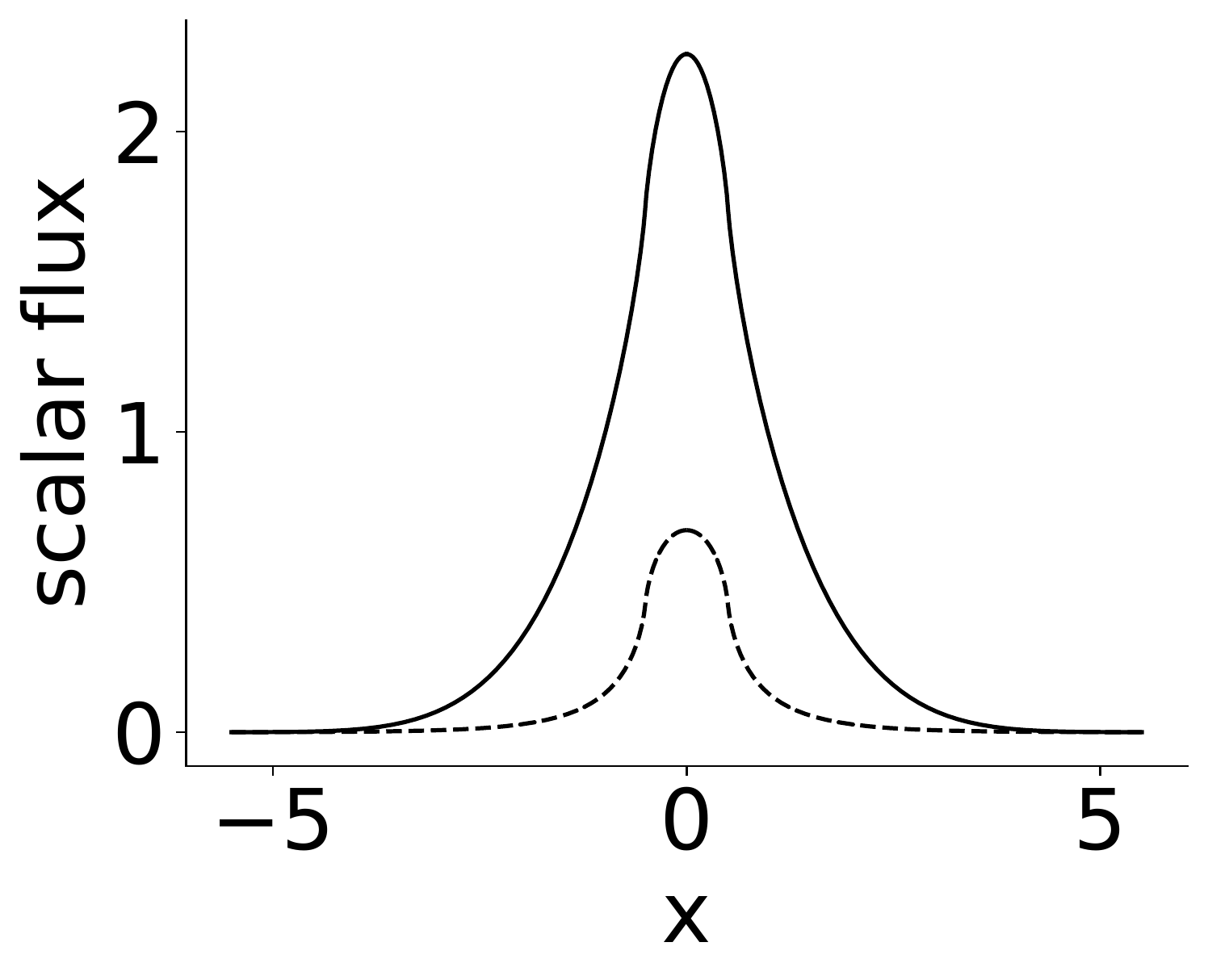}
         \caption{$t=5$}
         \label{fig:sq_s_5}
     \end{subfigure}
     \hfill
     \begin{subfigure}[b]{0.3\textwidth}
         \centering
         \includegraphics[width=\textwidth]{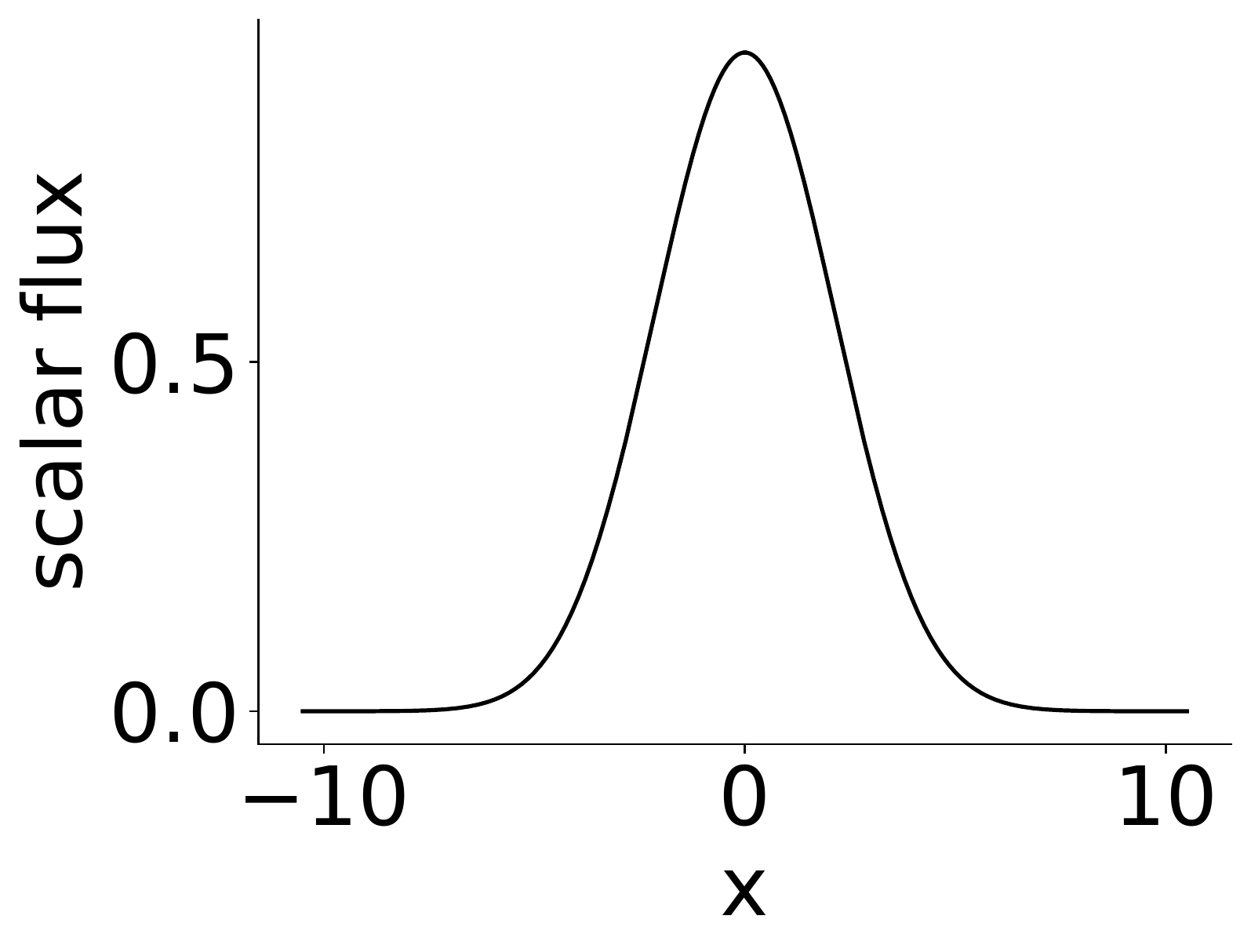}
         \caption{$t=10$}
         \label{fig:sq_s_10}
     \end{subfigure}
        \caption{Square source scalar flux solutions, $\phiu^\mathrm{ss} + \phic^\mathrm{ss}$, for $c=1$, $t_0 = 5$ and $x_0=0.5$ at several times; panels (a) and (b) also contain the uncollided scalar flux, $\phiu^\mathrm{ss}$, denoted by a dashed line.}
        \label{fig:sq_s}
\end{figure}

For this problem the solutions, as shown in Figure \ref{fig:sq_s} for $x_0=0.5$ and $t_0 = 5$, are smoother than the square pulse solutions shown previously. Also, because the source is one until $t=5$, there is a noticeable uncollided solution at that later time. 
\section{Gaussian source}
We next consider a Gaussian source with standard deviation $\sigma$ that is turned off at time $t_0$ where $S$ is given as
\begin{equation}\label{eq:gs}
    S(x,t) = \exp\left(\frac{-x^2}{\sigma^2}\right)\,\Theta(t_0 - t).
\end{equation}
Like the square pulse and the square source, the Gaussian source can be considered a superposition of Gaussian pulses, as Eq.~\eqref{eq:gs} is a superposition of pulses defined in Eq.~\eqref{eq:gp}. Therefore, the uncollided scalar flux solution can be found with a variable change from $t$ to $t-\tau$ in Eq.~\eqref{eq:gaussian_pulse_uncollided} and integration over the time that the source is on,
\begin{equation}\label{eq:gaussian_source_uncollided_1}
    \phiu^\mathrm{gs}(x,t) =  \int_0^{\mathrm{min}(t,t_0)}\!d\tau\,\sigma\,\sqrt{\pi }\, e^{-(t-\tau)} \,\frac{\text{erf}\left( \frac{t-\tau-x}{\sigma}\right)+\text{erf}\left(\frac{t-\tau+x}{\sigma}\right)}{4 (t-\tau)}.
\end{equation}
Solving this integral requires numerical integration. While this integral involves evaluating the integrand when $\tau = t$, the behavior of the error function allows the integrand to be well behaved.  Like the pulse source, this uncollided scalar flux is associated with a smooth angular flux solution.

\begin{figure}
     \centering
     \begin{subfigure}[b]{0.3\textwidth}
         \centering
         \includegraphics[width=\textwidth]{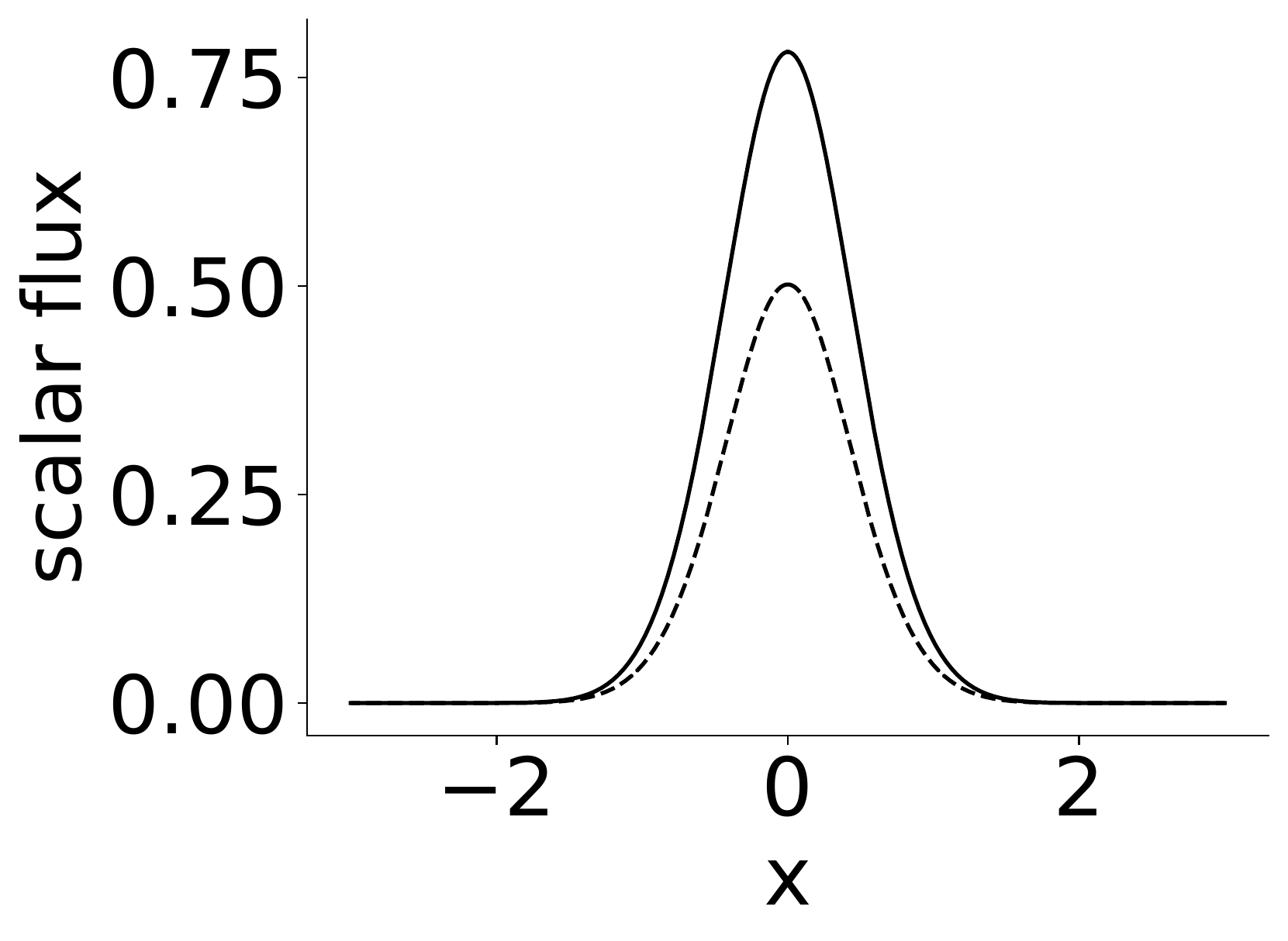}
         \caption{$t=1$}
         \label{fig:gs_s_1}
     \end{subfigure}
     \hfill
     \begin{subfigure}[b]{0.3\textwidth}
         \centering
         \includegraphics[width=\textwidth]{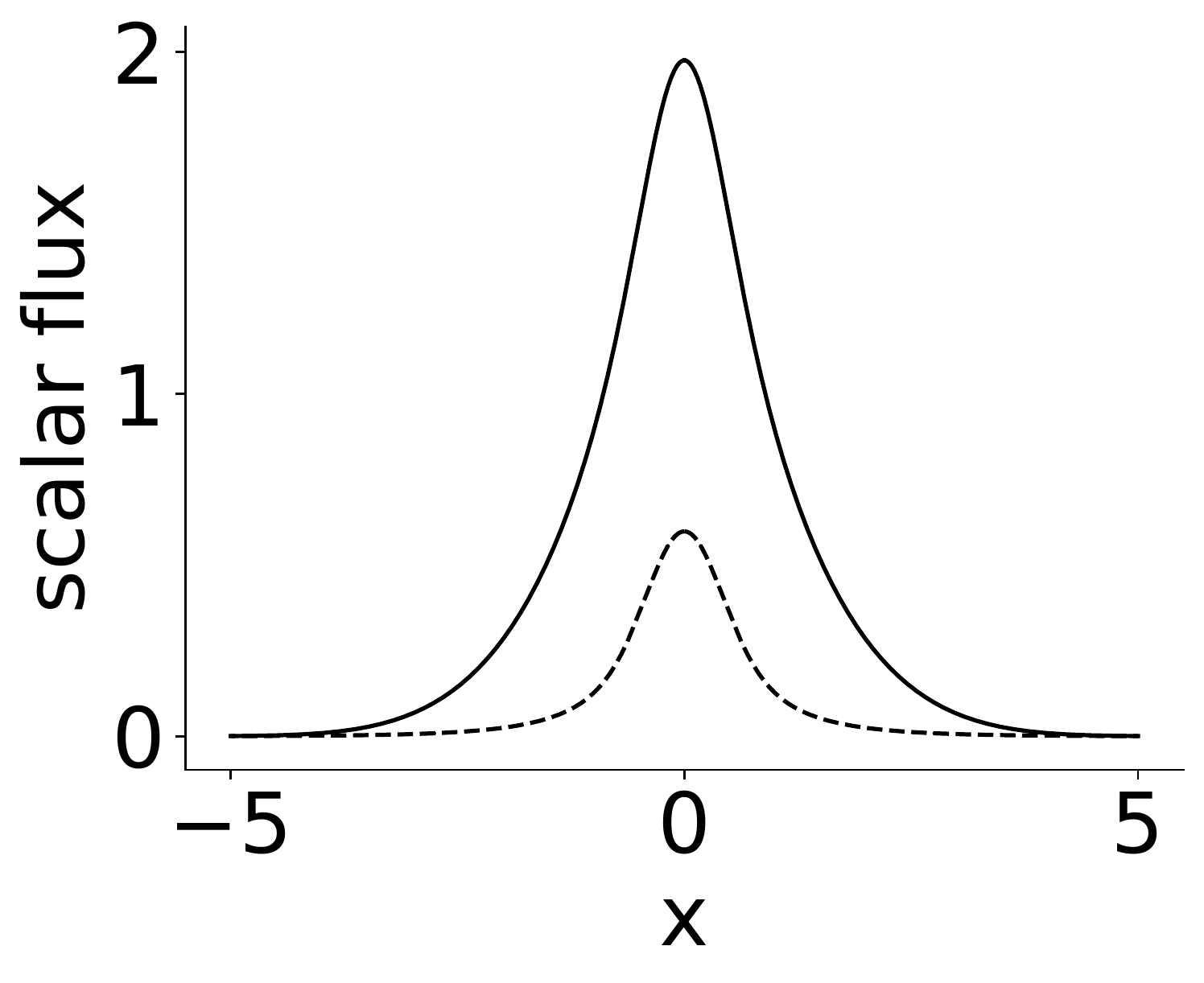}
         \caption{$t=5$}
         \label{fig:gs_s_5}
     \end{subfigure}
     \hfill
     \begin{subfigure}[b]{0.3\textwidth}
         \centering
         \includegraphics[width=\textwidth]{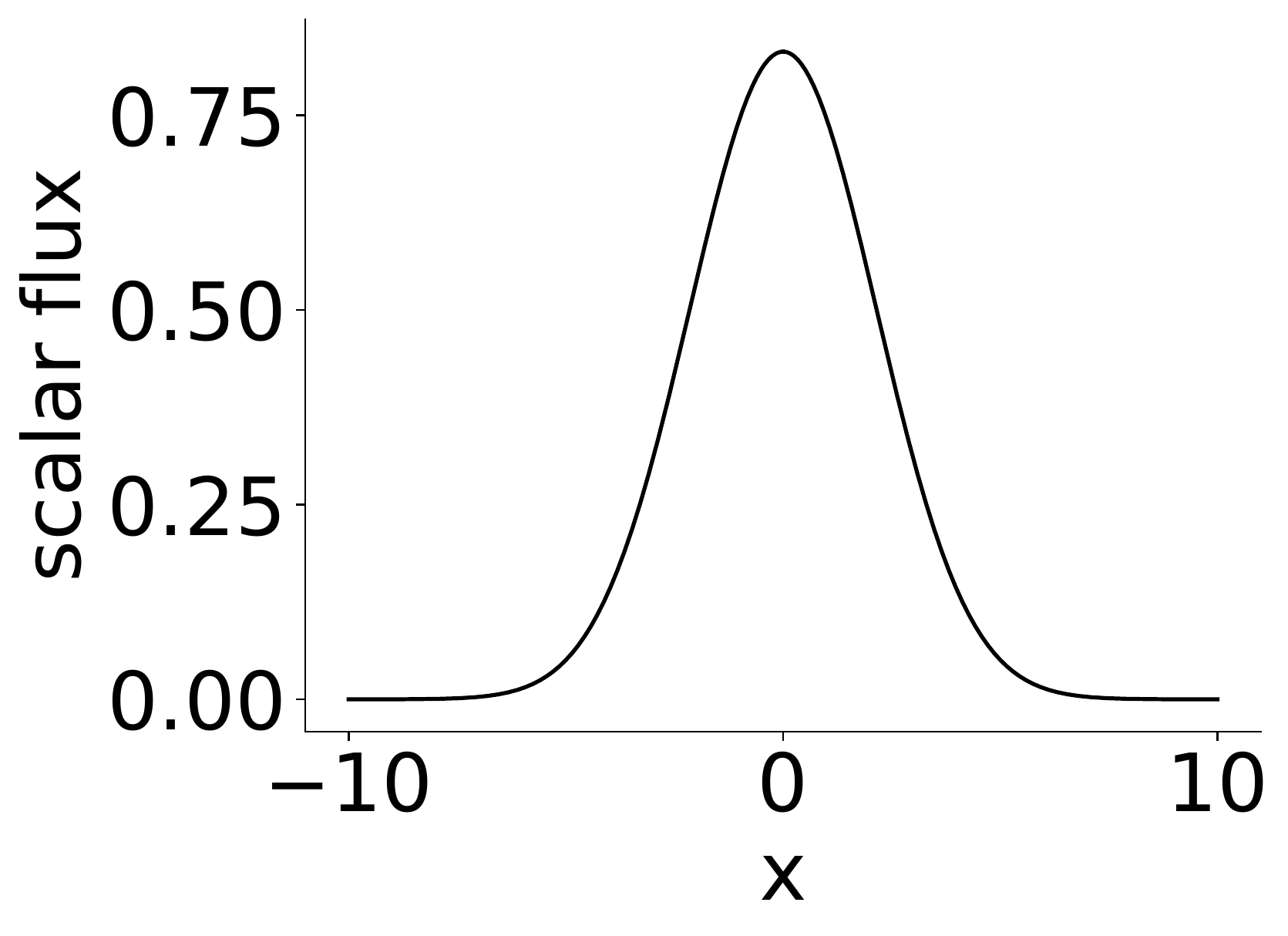}
         \caption{$t=10$}
         \label{fig:gs_s_10}
     \end{subfigure}
        \caption{Gaussian source scalar flux solutions, $\phiu^\mathrm{gs} + \phic^\mathrm{gs}$, for $c=1$, $t_0 = 5$ and $\sigma=0.5$ at several times; panels (a) and (b) also contain the uncollided scalar flux, $\phiu^\mathrm{gs}$, denoted by a dashed line.}
        \label{fig:gs_ss}
\end{figure}

To find the collided flux, we integrate Eq.~\eqref{eq:gauss_pulse_collided_1} over time,
\begin{equation}\label{eq:gaussian_pulse_collided_1}
    \phic^\mathrm{gs}(x,t) = \int_0^{\mathrm{min}(t,t_0)}\!d\tau\,\int^{\infty}_{-\infty}\!ds\,
    \exp\left({\frac{-s^2}{\sigma^2}}\right)\,F_2(x,s,t,\tau,u)\,\Theta\left(1-|\eta''|\right).
\end{equation}
Just as with the square source case (Section \ref{sec:sqs}), the integration orders for $\tau$ and $s$ are switched to find a better behaved integrand,
\begin{equation}\label{eq:gaussian_source_collided_1}
    \phic^\mathrm{gs}(x,t) = \int^{\infty}_{-\infty}\!ds\,\int^{\mathrm{min}(t,t_0)}_0\!d\tau\,
   \exp\left({\frac{-s^2}{\sigma^2}}\right)\,F_2(x,s,t,\tau,u)\Theta(1-|\eta''|).
\end{equation}
To find a more efficient integration interval over $s$, we  unravel $\eta''$ to get  $t-\tau-x \leq s \leq t-\tau+x$. With the floor of $\tau$ defined so that $\tau \geq 0$, new integration limits for $s$ are found and the most efficient form of the integral for the collided solution is,
\begin{equation}\label{eq:gaussian_source_collided}
    \phic^\mathrm{gs}(x,t) = \int^{x+t}_{x-t}\!ds\,\int^{\mathrm{min}(t,t_0,t-|x-s|)_+}_0\!d\tau\,
   \exp\left({\frac{-s^2}{\sigma^2}}\right)\,F_2(x,s,t,\tau,u),
\end{equation}
where $\mathrm{min}(\cdot)_+$ returns the minimum of its arguments or zero if the minimum is negative. 
Figure \ref{fig:gs_ss} shows the solutions for this problem with $\sigma = 0.5$ and $t_0 = 5$. These solutions resemble the square source solutions, especially at the later times. At $t=1$ the Gaussian source solution appears to be narrower than that for the square source solution.
\section{Cylindrical geometry}\label{sec:cyl_gaus}
\citet{ganapol} provides the  solution for cylindrical geometry where the source is an infinite line pulse aligned with the $z$ axis. This solution is obtained using the plane-to-point transform and integration of the point source over an infinite line.  The resulting uncollided expression is
\begin{equation}\label{eq:uncollided_line_source}
   \phiu^\mathrm{l}(r, t) = \frac{e^{-t}}{2\pi t^2}\frac{1}{\sqrt{1-\eta_p^2}}\Theta\left(1-\eta_p\right). 
\end{equation}
Here the superscript $\mathrm{l}$ denotes a line source, $r$ is the radial coordinate, and
\begin{equation}\label{eq:eta_cylindrical}
    \eta_p\equiv \frac{r}{t},
\end{equation} 
where the $p$ subscript stands for ``polar''.
The absolute value in the step function becomes irrelevant in this geometry and is discarded. Unlike the solutions presented thus far where there were no singularities apart from $t=0$, Eq.~\eqref{eq:uncollided_line_source} is singular as $\eta_p$ approaches one. The collided flux is found by integrating the collided flux for a point source, which Ganapol also provides,
\begin{equation}\label{eq:collided_line_source}
    \phic^\mathrm{l}(r,t) = \left[ 2t \int_0^{\sqrt{1-\eta_p^2}} \! d\omega\, \phic^\mathrm{pt}\left(t \sqrt{\eta_p^2 + \omega^2},t\right)\right]\Theta\left(1-\eta_p\right),
\end{equation}
where the point source collided flux is,
\begin{multline}
    \phic^\mathrm{pt}(r,t) = \Theta(1-\eta_p) \times\\ \left(\frac{e^{-t}}{4\pi r t^2} (ct) \log\left[\frac{1+\eta_p}{1-\eta_p}\right] + \frac{1}{2\pi} \frac{e^{-t}}{4 \pi r t^2} \left(\frac{ct}{2}\right)^2\left(1-\eta_p^2\right)\int_0^\pi \!du\, \sec^2\left(\frac{u}{2}\right) \mathrm{Re}\left[\left(\eta_p +i \tan \left(\frac{u}{2}\right)\right)\xi^3e^{\frac{ct}{2}\left(1-\eta_p^2\right)\xi}\right]\right),
\end{multline}
where the superscript $\mathrm{pt}$ is short for point, $\eta_p$ is given by Eq.~\eqref{eq:eta_cylindrical}, and $\xi$ by Eq.~\eqref{eq:xi}. The step function is redundant since it has been absorbed into the integration limits of Eq.~\eqref{eq:collided_line_source}.

\begin{figure}
     \centering
     \begin{subfigure}[b]{0.3\textwidth}
         \centering
         \includegraphics[width=\textwidth]{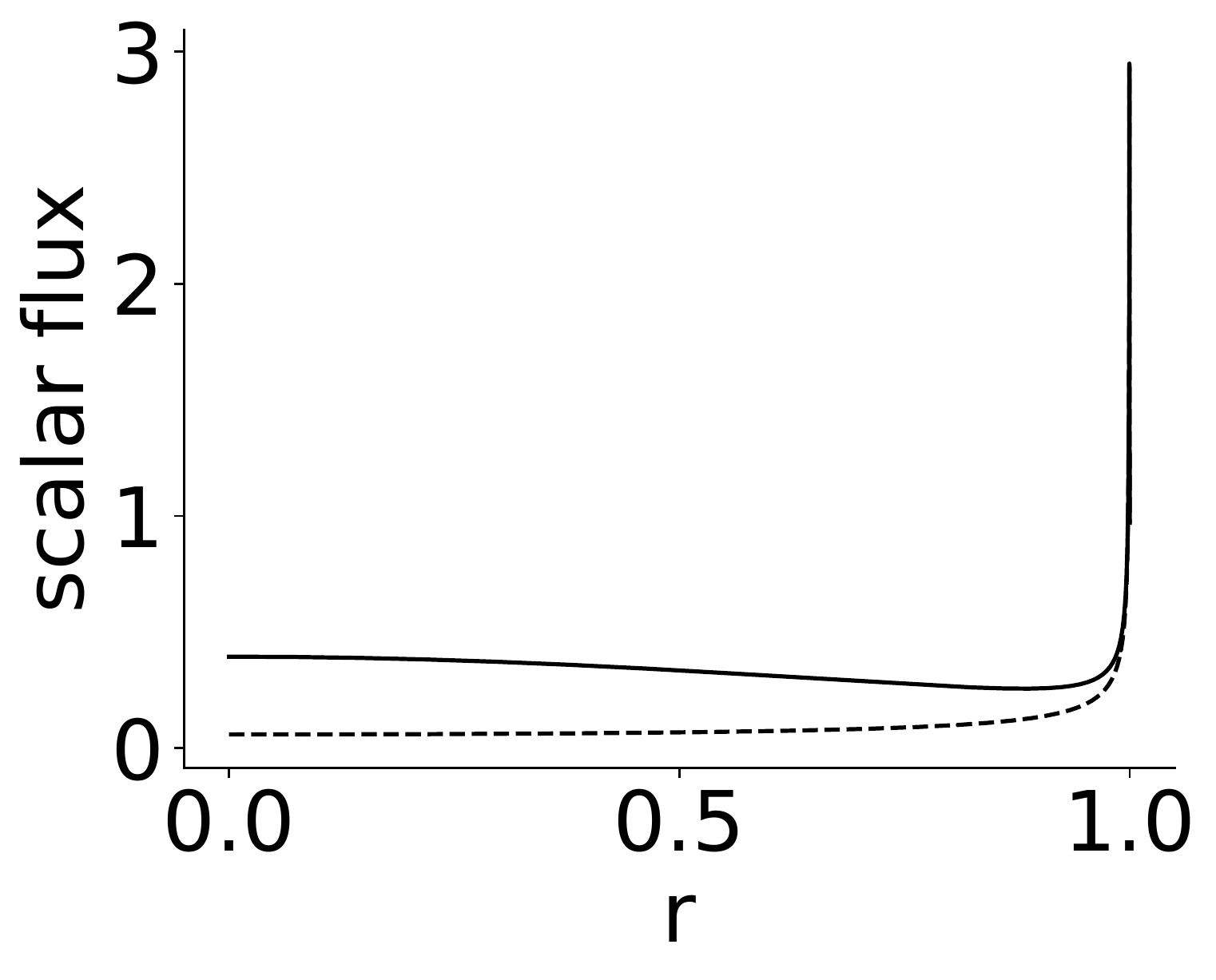}
         \caption{$t=1$}
         \label{fig:line_1}
     \end{subfigure}
     \hfill
     \begin{subfigure}[b]{0.3\textwidth}
         \centering
         \includegraphics[width=\textwidth]{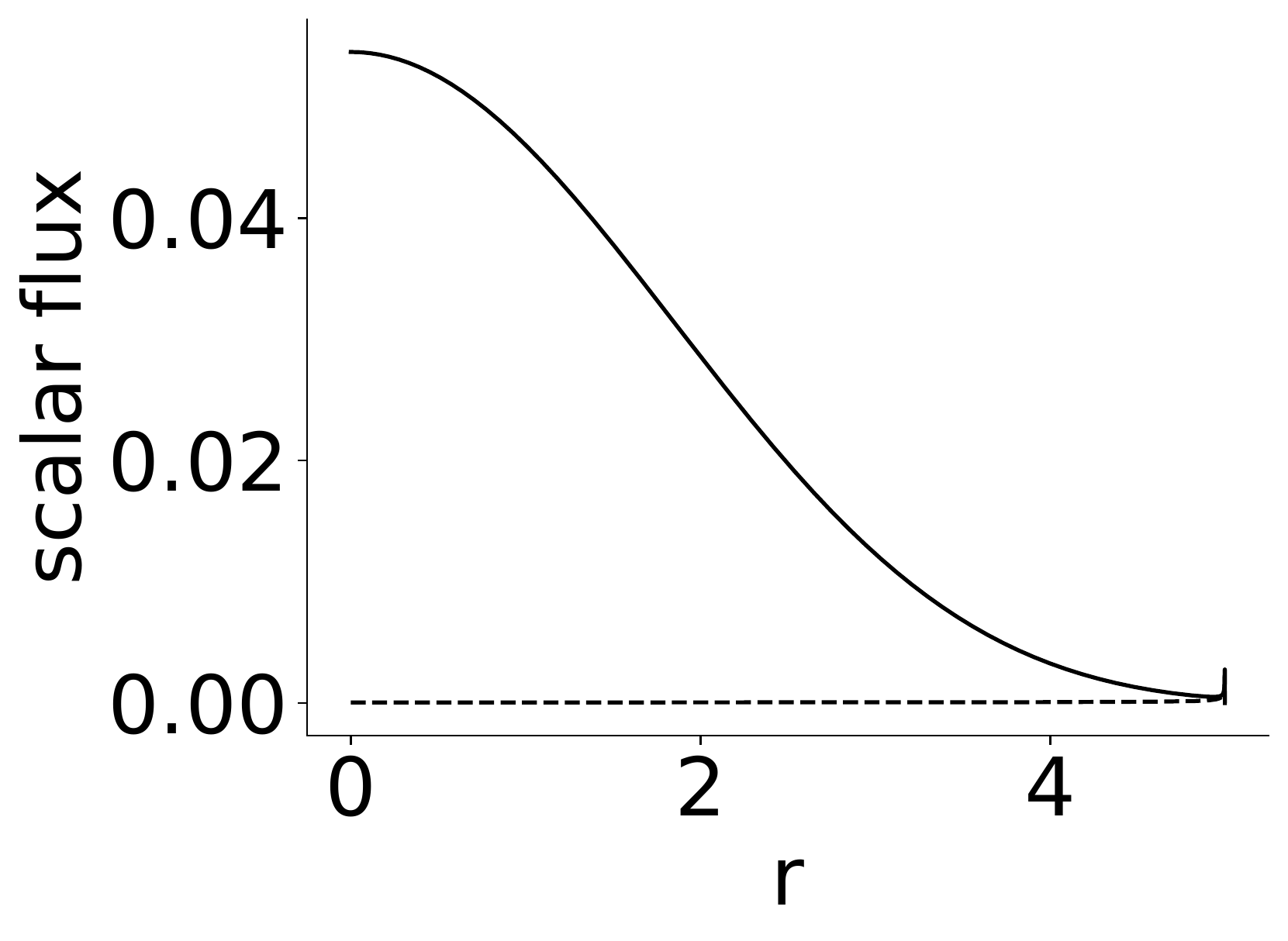}
         \caption{$t=5$}
         \label{fig:line_2}
     \end{subfigure}
     \hfill
     \begin{subfigure}[b]{0.3\textwidth}
         \centering
         \includegraphics[width=\textwidth]{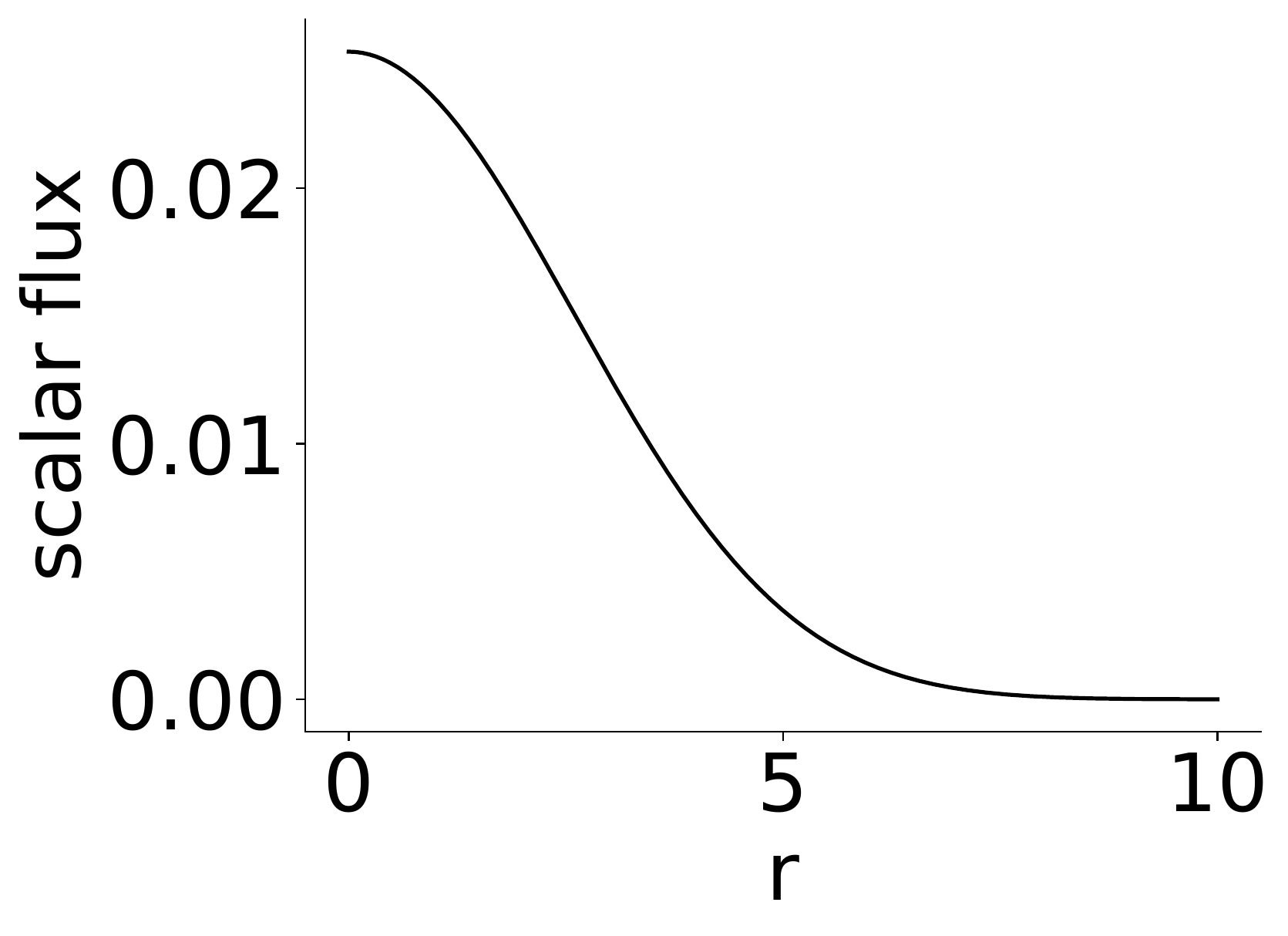}
         \caption{$t=10$}
         \label{fig:fig:g_s_10}
     \end{subfigure}
        \caption{Line pulse scalar flux solutions, $\phiu^\mathrm{l}+ \phic^\mathrm{l}$, for $c=1$, at several times; panels (a) and (b) also contain the uncollided scalar flux, $\phiu^\mathrm{l}$, denoted by a dashed line.}
        \label{fig:line}
\end{figure}

The line source solution is useful for verification of 2-D transport codes. However, as we can see from the solutions in Figure \ref{fig:line}, there is a singularity at the wavefront that is still present at $t=5$ on the scale of the figure.

\subsection{Gaussian pulse}
We consider an infinite cylindrical Gaussian pulse of standard deviation $\sigma$,
\begin{equation}
    S(r,t) = \exp\left(-\frac{r^2}{\sigma^2}\right)\delta(t). 
\end{equation}
For the uncollided flux, it is actually necessary to transform back into Cartesian coordinates since the integration in polar coordinates causes the integrand to be badly behaved. Using the relationship $r^2 = x^2 + y^2$, we introduce $r'$, 
\begin{equation}
    r'^2 = (x-s)^2 + (y-v)^2,
\end{equation}
Where $s$ and $v$ are dummy variables that are integrated over Cartesian space.
Defining a new $\eta$ for the Green's kernel, 
\begin{equation}\label{eq:eta_p_prime}
    \eta_p' = \frac{r'}{t}.
\end{equation}
Now the uncollided flux for a Gaussian pulse may be written in integral form,
\begin{equation}\label{eq:uncollided_gaussian_2d_1}
    \phiu^\mathrm{gp}(x, y, t) = \frac{e^{-t}}{2\pi t^2} \int^{\infty}_{-\infty}\!dv\, \int^{\infty}_{-\infty}\!ds\,  \frac{1}{\sqrt{1-\eta_p'^2}}\exp\left(-\frac{s^2 + v^2}{\sigma^2}\right)\Theta\left(1-\eta'_p\right). 
\end{equation}
The integrand of Eq.~\eqref{eq:uncollided_gaussian_2d_1} is still poorly behaved, but assimilating the step function into the integration limits of the integral over $s$ will cast it into a well behaved form. This is done by finding the roots of $s$ for the equation,
$\eta'_p = 1$. The expression for the uncollided flux becomes, 
\begin{equation}\label{eq:uncollided_gaussian_2d_final}
       \phiu^\mathrm{gp}(x, y, t) = \frac{e^{-t}}{2\pi t^2} \int^{\infty}_{-\infty}\!dv\, \int^{s_b}_{s_a}\!ds\,  \frac{1}{\sqrt{1-\eta_p'^2}}\exp\left(-\frac{s^2 + v^2}{\sigma^2}\right), 
\end{equation}
where the integration limits over $s$ are, 
\begin{equation}
    s_a = 
    x-\sqrt{[t^2-v^2+2 v y-y^2]_+} 
\end{equation}
\begin{equation}
    s_b = 
    x+\sqrt{[t^2-v^2+2 v y-y^2]_+}
\end{equation}
where $[\cdot]_+$ returns the positive part of its argument.
Since the solution is symmetric about the pole, it is not necessary to integrate Eq.~\eqref{eq:uncollided_gaussian_2d_final} over a two dimensional domain. We can choose $y=0$ and find the uncollided solution as a function of $r$.
\begin{equation}\label{eq:uncollided_gaussian_2d_r}
    \phiu^\mathrm{gp}(r,t) = \frac{e^{-t}}{2\pi t^2} \int^{\infty}_{-\infty}\!dv\, \int^{s_b}_{s_a}\!ds\,  \frac{1}{\sqrt{1-\eta_p''^2}}\exp\left(-\frac{s^2 + v^2}{\sigma^2}\right), 
\end{equation}
where
\begin{equation}
    \eta_p'' = \frac{\sqrt{(r-s)^2+v^2}}{t}.
\end{equation}

The expression for the collided flux is better behaved in polar coordinates, where the  variables over which the Green's kernel is integrated become $\rho$ and $\theta'$. With this transformation, we define a new radius in polar coordinates,
\begin{equation}\label{eq:r_prime}
    r' = \sqrt{(r \cos(\theta) - \rho \cos(\theta'))^2 + (r \sin(\theta) - \rho \sin(\theta'))^2}.
\end{equation}
$\eta_p'$ is given by Eq.~\eqref{eq:eta_p_prime} with Eq.~\eqref{eq:r_prime} as $r'$.
Since the solution will be symmetric about $r$, the value of angular coordinate $\theta$ is arbitrary. However, to properly integrate the solution kernel, the angular coordinate, $\theta'$, must be independent from $\theta$. Therefore, the collided flux requires integration over angle and radius,  
\begin{equation}\label{eq:collided_gaussian_2d_1}
    \phic^\mathrm{gp}(r,\theta, t) = 2t \int^{2\pi}_0\!d\theta'\,\int^{\infty}_0\!d\rho\,\rho\exp\left(-\frac{\rho^2}{\sigma^2}\right)\left[ \int_0^{\sqrt{1-\eta_p'^2}} \! d\omega\, \phic^\mathrm{pt}\left(t \sqrt{\eta_p'^2 + \omega^2},t\right)\right]\Theta\left(1-\eta_p'\right).
\end{equation}
Evaluating Eq.~\eqref{eq:collided_gaussian_2d_1} requires four integrals over a difficult integrand. To simplify the integral, we first recast it in a more explicit form where the step function from the point source is brought out of the function,
\begin{multline}
      \phic^\mathrm{gp}(r,\theta, t)  = 2t \int^{2\pi}_0\!d\theta'\,\int^{\infty}_0\!d\rho\,\int^{\sqrt{1-\eta_p'^2}}_0\!d\omega\,\int^\pi_0\!du \,Q(\rho)\, \rho\, F^\mathrm{pt}_2\left(t \sqrt{\eta_p'^2 + \omega^2},t\right) \Theta(1-\eta'_p) + \\
       2t \int^{2\pi}_0\!d\theta'\,\int^{\infty}_0\!d\rho\,\int^{\sqrt{1-\eta_p'^2}}_0\!d\omega\,Q(\rho)\, \rho\, F^\mathrm{pt}_1\left(t \sqrt{\eta_p'^2 + \omega^2},t\right) \Theta(1-\eta_p'),
\end{multline}
Where $F_1^\mathrm{pt}$ is the first collided kernel for a point source without the step function,
\begin{equation}
    F^\mathrm{pt}_1(r,t) = \frac{e^{-t}}{4\pi r t^2} (ct) \log\left[\frac{1+\eta_p}{1-\eta_p}\right],
\end{equation}
and $F_2^\mathrm{pt}$ is the integrand for the second on to infinite collided solution for a point source without the step function,
\begin{equation}
    F_2^\mathrm{pt}(r,t) =  \frac{1}{2\pi} \frac{e^{-t}}{4 \pi r t^2} \left(\frac{ct}{2}\right)^2\left(1-\eta_p^2\right) \sec^2\left(\frac{u}{2}\right) \mathrm{Re}\left[\left(\eta_p +i \tan \left(\frac{u}{2}\right)\right)\xi^3e^{\frac{ct}{2}\left(1-\eta_p^2\right)\xi}\right],
\end{equation}
where $\eta_p$ is given by Eq.~\eqref{eq:eta_cylindrical}.
The source, $Q(\rho)$ is 
\begin{equation}
    Q(\rho) = \exp\left(-\frac{\rho^2}{\sigma^2}\right).
\end{equation}
The step function, $\Theta(1-\eta_p')$, causes the effective integration domain to be erratic and the integrand to be badly behaved. As with the uncollided case, solving $\eta_p' = 1$ gives the upper and lower bounds of the integration for $\rho$ as dictated by the step function. Now a convergent form of collided flux can be found
\begin{multline}
      \phic^\mathrm{gp}(r,\theta, t)  = 2t \int^{2\pi}_0\!d\theta'\,\int^{\rho_b}_{\rho_a}\!d\rho\,\int^{\sqrt{1-\eta_p'^2}}_0\!d\omega\,\int^\pi_0\!du \,Q(\rho)\, \rho\, F^\mathrm{pt}_2\left(t \sqrt{\eta_p'^2 + \omega^2},t\right) \Theta(1-\eta'_p) + \\
       2t \int^{2\pi}_0\!d\theta'\,\int^{\rho_b}_{\rho_a}\!d\rho\,\int^{\sqrt{1-\eta_p'^2}}_0\!d\omega\,Q(\rho)\, \rho\, F^\mathrm{pt}_1\left(t \sqrt{\eta_p'^2 + \omega^2},t\right) \Theta(1-\eta_p'),
\end{multline}
where
\begin{equation}
    \rho_a = r \cos (\theta -\theta' )-\sqrt{\left[\frac{r^2 \cos (2 (\theta -\theta' ))-r^2}{2} +t^2\right]_+},
\end{equation}
\begin{equation}
    \rho_b = r \cos (\theta -\theta' ) + \sqrt{\left[\frac{r^2 \cos (2 (\theta -\theta' ))-r^2}{2} +t^2\right]_+},
\end{equation}
\begin{equation}\label{eq:eta_polar}
    \eta'_p = \frac{\sqrt{(r \cos(\theta) - \rho \cos(\theta'))^2 + (r \sin(\theta) - \rho \sin(\theta'))^2}}{t}.
\end{equation}

\begin{figure}
     \centering
     \begin{subfigure}[b]{0.3\textwidth}
         \centering
         \includegraphics[width=\textwidth]{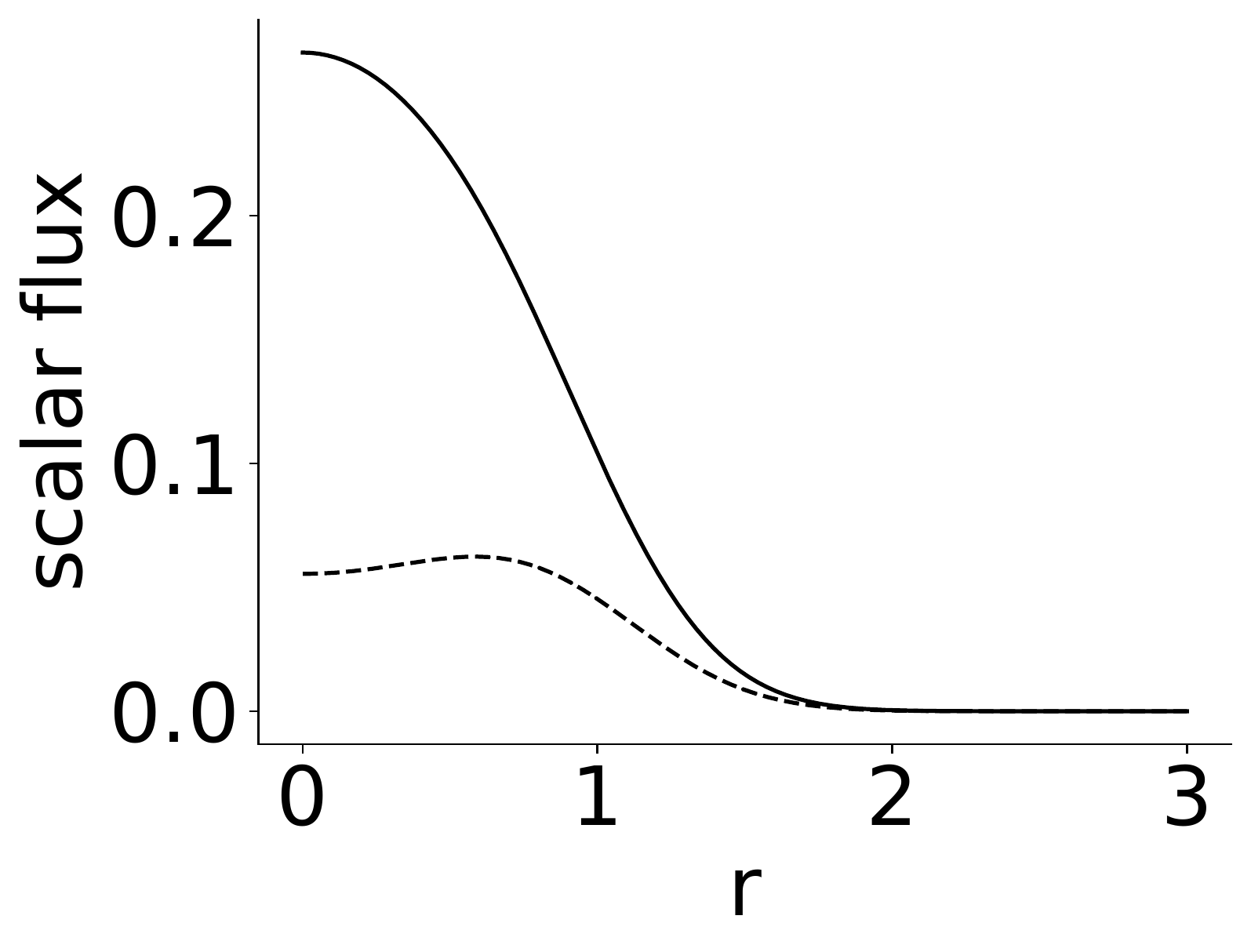}
         \caption{$t=1$}
         \label{fig:gs_2d_1}
     \end{subfigure}
     \hfill
     \begin{subfigure}[b]{0.3\textwidth}
         \centering
         \includegraphics[width=\textwidth]{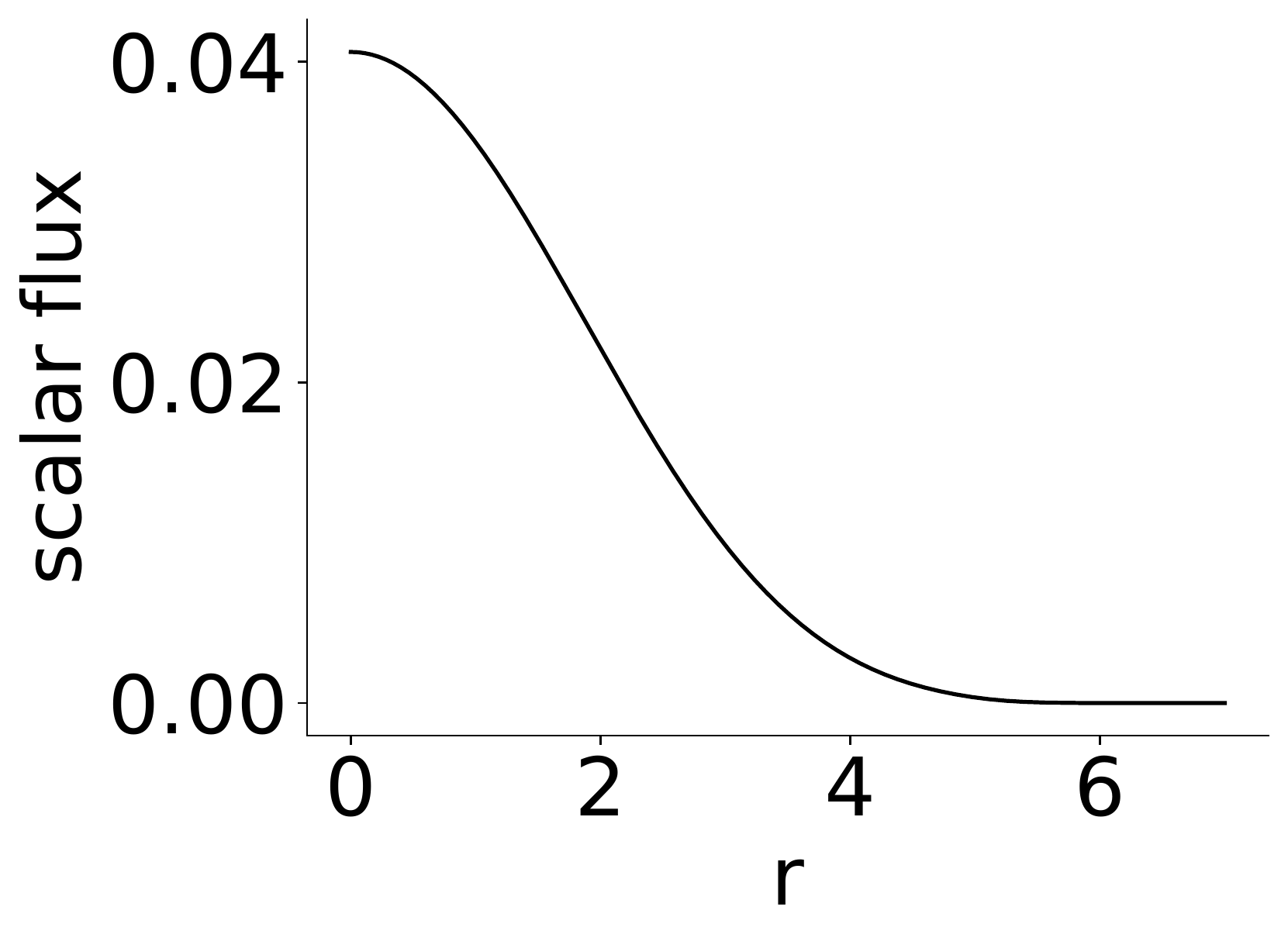}
         \caption{$t=5$}
         \label{fig:gs_2d_5}
     \end{subfigure}
     \hfill
     \begin{subfigure}[b]{0.3\textwidth}
         \centering
         \includegraphics[width=\textwidth]{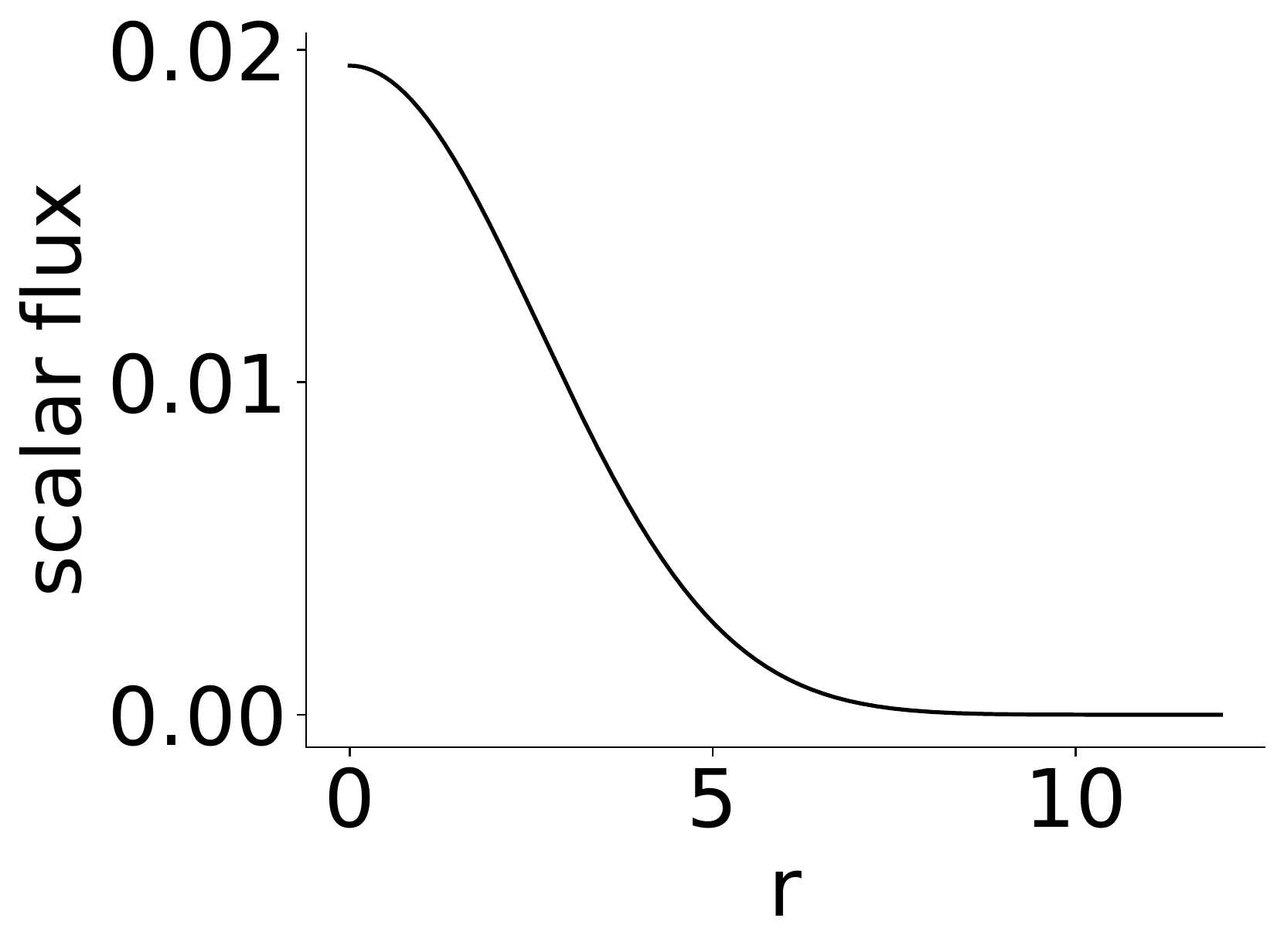}
         \caption{$t=10$}
         \label{fig:gs_2d_10}
     \end{subfigure}
        \caption{Cylindrical Gaussian source scalar flux solutions, $\phiu^\mathrm{gs} + \phic^\mathrm{gs}$, for $c=1$, $t_0 = 5$ and $\sigma=0.5$ at several times; panel (a) also contains the uncollided scalar flux, $\phiu^\mathrm{gs}$, denoted by a dashed line.}
        \label{fig:gs_sc}
\end{figure}

This solution with $\sigma = 0.5$ and $t_0=5$ is shown in Figure \ref{fig:gs_sc}. Integrating the source over a finite spatial range leads to a removal of the singularity in the solution at the wavefront. This leads to a smooth solution at all times.

\section{Conclusion}
We have presented the uncollided and collided solutions for several different source shapes that are either  pulsed or turned-on for a finite time. In many cases we were able to produce a simple formula for the uncollided solution so that could be used as a source in a computer code for verification purposes in a similar manner to MMS.  We believe that these solutions will be useful for code verification for a variety of transport codes.

All of the solutions presented were checked for systematic errors with an independent numerical solver that discretized the transport equation.  A package that evaluates all of the presented solutions along with precomputed numerical results for $t=1,\,5,\,\mathrm{and}\,10$ are available on GitHub\footnote{{github.com/wbennett39/transport\textunderscore benchmarks}.}. The code can be used to modify the parameters in the solutions to produce different benchmarks.

\section*{Acknowledgements}
The authors would like to thank Minwoo Shin for providing independent numerical solutions to the cylindrical Gaussian pulse (Sec. \ref{sec:cyl_gaus}).
This work was supported by a Los Alamos National Laboratory under contract \#599008, ``Verification for Radiation Hydrodynamics Simulations''.

\bibliographystyle{tfcad}
\bibliography{ref}

\end{document}